\documentclass[12pt]{article}
\usepackage[a4paper, margin=2cm]{geometry}
\usepackage{natbib}
\usepackage{amsmath}
\usepackage[final]{changes}
\usepackage{amsfonts}
\usepackage{fancyvrb}
\usepackage{caption}
\usepackage{pbox}
\usepackage{float}
\usepackage{multirow}
\usepackage{hyperref}
\usepackage{authblk}
\usepackage{tablefootnote}
\usepackage{setspace}
\usepackage{tabularx}
\usepackage{mathtools}
\usepackage{adjustbox}
\usepackage{multirow}
\usepackage[flushleft]{threeparttable}
\usepackage{listings}
\usepackage{subcaption}
\usepackage[title]{appendix}
\usepackage{amssymb}
\usepackage{enumitem}
\usepackage{bbm}
\usepackage{verbatim}
\usepackage{longtable}
\newcommand{\pkg}[1]{{\normalfont\fontseries{b}\selectfont #1}}
\let\proglang=\textsf
\let\code=\texttt
\definechangesauthor[name={TMOTTM},color=red]{TT}
\title{Using saturated count models for user-friendly synthesis of categorical data}
\author[$\dagger$]{James Jackson}
\author[$\star$]{Robin Mitra}
\author[$\dagger$]{Brian Francis}
\author[$\ddag$]{Iain Dove}
\affil[$\dagger$]{Lancaster University, Lancaster, UK}
\affil[$\star$]{Cardiff University, Cardiff, UK}
\date{}
\affil[$\ddag$]{Office for National Statistics, Titchfield, UK}
\begin{document}
\maketitle
\begin{abstract}
Over the past three decades, synthetic data methods for statistical disclosure control have continually {evolved}, but mainly within the domain of survey data sets. {There are c}ertain characteristics of administrative databases, {such as their size, which} present challenges from a synthesis perspective and require special attention. This paper, through the fitting of saturated {count} models, presents a {synthesis method that is suitable for administrative databases. It is tuned by two parameters, $\sigma$ and $\alpha$. The method allows large categorical data sets to be synthesized quickly and allows risk and utility metrics to be satisfied \textit{a priori}, that is, prior to synthetic data generation}.  The paper explores how the flexibility afforded by two-parameter count models (the negative binomial and Poisson-inverse Gaussian) can be utilised to protect respondents' - especially uniques' - privacy in synthetic data. Finally, an empirical example is carried out through the synthesis of a database which can be viewed as a good {substitute} to the English School Census.
\end{abstract}
\section{Introduction}
There is an increasing demand for data to be made available to researchers. {T}his is coupled{, however,} with greater demands, both legal and ethical, on the protection of personal data. Consequently, new data sources, and innovative ways of protecting data, are required. \par 
Administrative databases originate because organisations need to hold individuals' information for their day-to-day running. They have been largely under-explored as a potential data source, but {can} contain vast amounts of information, sometimes for an entire population. For this reason, administrative data are being used to enhance the UK census; and the UK's National Statistician, Professor Sir Ian Diamond, has gone further and recommended that  administrative data can help to replace future censuses \citep{HMGovernment2018}; censuses are, after all, notoriously expensive.\par 
Often administrative databases are hidden away with access limited to, for example, staff at government departments. When data are made available to researchers, it is usually via controlled environments, such as the Secure Research Service (SRS), a facility within the Office for National Statistics (ONS). To access data within the SRS, researchers have to undertake formal training, submit project applications and then, depending on the data’s confidentiality, process the data in safe rooms. While this procedure is{, of course,} necessary, it can be time-consuming {and} may deter researchers. Alternatively, in some cases it is possible to release open data, which is when data are released into the public domain for anyone to access. \par Protecting data confidentiality is the main priority when disseminating any individual-level data set. In the UK, the General Data Protection Regulation (GDPR) means that businesses and organisations are legally obliged to adhere to certain standards with regards to anonymisation when processing personal data \citep{ICO2020}. Anonymisation is achieved through applying statistical disclosure control (SDC) methods; {see \cite{Duncan2011}, \cite{Hundepool2012} and \cite{Templ2017} for a thorough review of such methods}. \par {Particularly stringent anonymisation techniques are required for administrative data - even if the data are made available in a secure environment. Administrative data are particularly sensitive, as individuals do not explicitly supply their own information as they would when responding to a survey. The disclosure of sensitive information would adversely affect the reputations of those involved in processing the data, which may ultimately affect their ability to collect and process data.}\par   The use of synthetic data \citep{Rubin1993, Little1993} to protect privacy continues to attract attention. Whereas traditional methods typically either perturb or suppress the original data until a satisfactory level of protection is attained, synthetic data methods involve constructing new data sets by simulating from models fitted to the original data. \par In 1993 the \textit{Journal of Official Statistics} published a special issue on data confidentiality. Two contributions therein planted the seed for the growth of synthetic data sets. \cite{Rubin1993} proposed to multiply impute values for those individuals in the population who were not sampled in the original data, and release simple random samples; these are now widely known as fully synthetic data sets \citep{Raghunathan2003}. \cite{Little1993} proposed a similar idea whereby only certain values in the data are replaced; these are now widely known as partially synthetic data sets \citep{Reiter2003}. The idea is that the synthetic data can replace the original data and provide analysts with {similar} inferences to those that would have been obtained had the analysis been performed on the original data. \par {As synthetic data are inherently artificial, disclosure risks are minimal. In statistical databases, two types of disclosure can be considered: \textit{re-identification} and \textit{attribute disclosure} (there is also \textit{inferential disclosure}, though this is closely related to attribute disclosure). Re-identification is when an {attacker} de-anonymises an anonymised record, and so identifies an individual in the data. Attribute disclosure is when an attacker can precisely estimate an individual's sensitive values, without necessarily being able to identify them in the data set.} Re-identification becomes meaningless in fully synthetic data because individuals in the synthetic data do not directly pertain to individuals in the original data. This is especially beneficial for administrative {databases, which} may hold information for an entire population, so there is no natural protection through sampling uncertainty, that is, uncertainty as to whether an individual was actually included in the original data. {However, a synthesizer needs to guard against attribute disclosure because, through the synthetic data's release, an attacker can potentially deduce certain sensitive information. The correct attribution probability (CAP) metric, see \cite{Taub2018}, seeks to measure the risk of attribute disclosure in synthetic data.} {For more about disclosure risk in microdata more generally see \cite{Duncan1989}, or \cite{Hu2019} for risk in synthetic data specifically.} \par
The risk-utility trade-off{, proposed by \cite{Duncan2001},} is inherent in all data dissemination: high utility typically comes at the expense of high risk of disclosure. The ONS devised a spectrum \citep{Bates2019} as a way of classifying the position of synthetic data with respect to this trade-off. At one end of the spectrum are ``structural'' synthetic data sets where only the original data {set}'s general structure {is} preserved, such as variable names; these could be employed as test data, that is, data on which researchers can first run their analyses to identify, for example, any issues with code, before repeating their analyses on the original data. At the other end of the spectrum are ``replica'' synthetic data sets, which are designed to be analysed in place of the original data. {The method introduced in this paper can be used to produce synthetic data anywhere on this spectrum.} \par {The synthetic data literature stems from the literature for the multiple imputation of} missing data. {An} appealing feature of multiple imputation {is} that the burden of imputing missing values f{alls} on the imputer - a trained statistical modeller - rather than the analyst, who may be less well-versed in {s}tatistics. This philosophy carried over into the development of synthesis methods, which increasingly utilise complex computational techniques, necessitating specific non-trivial modelling decisions that require specialist training {and often large amounts of recorded central processing unit (CPU)} time to implement effectively. However, this is coupled with data-holders (many of whom would not be trained in these advanced statistical methods) taking greater interest in producing their own synthetic data and thus retaining greater control over the synthesis process. There is{, therefore,} growing appeal in developing synthesis methods that ease the burden on the synthesizer while still generating appropriate synthetic data that satisfy data-holders' requirements. This is a key motivating principle that underpins our proposed methodology. \par {The synthesis method presented in this paper provides a quick way to synthesize large categorical data sets. Overdispersed, saturated synthesis models are used to: (i) overcome constraints in model fitting, (ii) preserve relationships and (iii) allow risk and utility metrics to be satisfied in an \textit{a priori} fashion. The method takes a drain-and-inject approach to synthesis: uncertainty from modelling is drained away and, instead, uncertainty is injected where it is most needed, which is to protect the records at greatest risk of disclosure.}\par
{Through tuning two parameters (introduced later as $\sigma$ and $\alpha$), the synthesizer can immediately generate synthetic data with different levels of risk and utility. In this way, the method shares a trait with differentially private mechanisms \citep{Dwork2006}, which also preserve privacy through tuning a parameter (usually denoted by $\epsilon$). Methods tuned by a parameter allow noise to be applied as appropriate: for example, when the privacy budget is high, the mechanism can easily be adjusted to reduce noise and hence risk (and vice versa). This has the potential to allow risk to be considered in a more formal way. }
 \par This paper shows that synthetic data sets are a viable option for safely making information held in administrative databases available to researchers, which will hopefully stimulate further interest and development. The paper is structured as follows: Section 2 reviews existing synthesis methods, {with the focus on} categorical variables, {and} considers particular challenges faced when synthesizing large administrative databases. Section {3} introduces this paper's contribution to the field: the $(\sigma, \alpha)$-synthesis method, which uses saturated models. Section {4} presents an empirical illustration: the synthesis of a database which {can be viewed as a substitute to} the English Schools Census. Section 5 gives some concluding remarks.
\section{Methods of synthetic data generation {for categorical data}}
{Typically data sets, including administrative databases, naturally take a microdata format, where the individuals form the rows and the variables the columns.} Suppose a synthesizer wishes to synthesize a microdata set $Y=(Y_1, Y_2, \hdots, Y_p)$ comprising $n$ {individuals} completely observed over $p$ variables, {so that the data form a $n \times p$ matrix.} The first step in synthetic data generation involves modelling the joint multivariate distribution of {this} data {$Y$}{. F}or categorical data, this can be carried out at either the microdata level or at the aggregated (tabular) level. 
\subsection{{Synthesizing microdata}}
\cite{Drechsler2011} {describes two broad methods for generating synthetic microdata: conditional and joint approaches.} The conditional approaches model the original data through a product of conditional univariate models, that is,
\begin{align}
p(Y_1, Y_2, \hdots, Y_p) = p(Y_1) \prod_{j=2}^p p(Y_{j} \mid Y_{j-1} \hdots, Y_2, Y_1).
\end{align}
Separate models can then be specified for each variables. This approach is flexible in the sense that it can {deal with data sets that are comprised of} different variable types{. For example, normal linear regression can be used to model continuous variables and} multinomial logistic regression models can be {used} to model categorical variables. \par {Joint modelling approaches, on the other hand, specify a multivariate model for the entire data set. For example, if all variables are continuous it may be possible to fit a multivariate normal distribution. However, this can be difficult to implement in practice, especially using parametric models.}
\subsubsection{{Conditional approaches}} \label{conditional}
{As described above, conditional approaches begin by modelling the first variable; then by  modelling the second variable conditional on the first; the third conditional on the first and second; and so on, up to the $p$th variable, which is conditional on all other $p-1$ variables.} Multinomial logistic regression models are an obvious choice for modelling categorical variables. Although not strictly a generalized linear model (GLM) 
owing to the multivariate response, the multinomial logistic regression model can be viewed as an extension to the (binary) logistic regression model to the case where the response has three or more categories. \par When there are many regression coefficients, {however}, fitting these models is beyond the capabilities of the algorithms used in standard statistical software. {Nevertheless}, when the model's covariates are categorical, the time taken to obtain the regression coefficients' maximum likelihood estimates can be substantially reduced by utilising the Poisson-multinomial equivalence. Every multinomial model has a corresponding Poisson log-linear model; {see, for instance,} \cite{Lang1996}. {Fitting the corresponding Poisson log-linear model} allows the iterative proportional fitting (IPF) algorithm \citep{Deming1940} to be used, which provides a quick-and-easy route to obtain the model's fitted values (the expected counts). {\cite{Skinner2008} used IPF to fit log-linear models when estimating disclosure risk in microdata.} However, a downside of IPF is that, while expected counts are obtained, regression coefficients' {estimates and standard errors} are not. This has implications for generating fully synthetic data as described by \cite{Rubin1993}, where parameters' estimates and standard errors are intrinsic to deriving parameters' posterior distributions.
\par Classification and regression trees (CART), {which were developed by \cite{Breiman1984} and} can be viewed as a non-parametric analogue to the GLM{, were} considered as a method to generate partially synthetic data by \cite{Reiter2005}. CART generates synthetic data sequentially, by growing a tree for each variable, conditional on all other variables in the data. Its appeal has increased with the \proglang{R} package \pkg{synthpop} \citep{Nowok2016}, for which CART is the default synthesis method. \par {Over time, CART logically led to} the use of random forests for synthesis \citep{Caiola2010}. Also developed by \cite{Breiman2001}, random forests grow multiple trees per variable. \cite{Drechsler2011a} demonstrated the effectiveness of these non-parametric tree-based methods for synthesis, relative to {parametric} approaches.  \par 
\subsubsection{{Joint modelling approaches}}
{The non-parametric latent class model (NPLCM) \citep{Dunson2009}, which is a Dirichlet process mixture of products of multinomial (DPMPM) distributions, can be used to generate synthetic categorical data \citep{Hu2014,Manrique-Vallier2014,Manrique-Vallier2018}}. {As with the latent class model given by \cite{Goodman1974}, the model assumes the existence of $F\geq1$ latent classes and introduces a set of latent class probabilities $\pi_1, \hdots, \pi_F \;(\sum_{i=1}^F\pi_i=1)$, where $\pi_i$ is the probability that an individual belongs to latent class $i$. Then, within each latent class, a specific multinomial distribution can be fit, resulting in a flexible mixture model that can correspond to many different distributions.} The model has a fully Bayesian specification, {so} Markov-Chain Monte Carlo (MCMC) methods are required to obtain {samples from the posterior distribution}. {This} can be carried out via the \proglang{R} package \pkg{NPBayesImputeCat}  \citep{Hu2021}. \par {In addition, machine learning techniques are becoming an increasingly popular area of research in relation to synthetic data, such as the use of generative adversarial networks (GANs) \citep{Kaloskampis2019}.} 
\subsection{{Synthesizing aggregated counts}}
{When all variables are categorical, there are a finite number of possible observations that any individual can observe, which is determined by the number of category combinations across variables. This means that, without loss of information, the data can be expressed as a multi-dimensional contingency table, where counts give the number of times the combinations of categories are observed. In general, if there are $p$ categorical variables with $l_1, \hdots, l_p$ categories, respectively, then the data can be cross-tabulated and expressed as a table with $K=l_1 \times \hdots \times l_p$ cell counts, where each count gives the number of individuals who belong to a particular cell.} \par
The data can {then} be synthesized at the aggregated level, rather than the individual level, by modelling these counts. 
\subsubsection{The Poisson log-linear model} 
The {counts in the multi-way table} can be modelled by a Poisson log-linear model, which assumes that the counts are independent and Poisson distributed. The model has a representation as a generalized linear model, in which it is parameterized by an intercept term, main effects and interaction effects. The interactions pertain to associations between variables; whenever an interaction is set to zero, independence is assumed between those variables. \par {The $i$th synthetic cell count $(i=1,\hdots,K)$ of the multi-way table $f_i^\text{syn}$, is modelled as follows:}
\begin{align}
f_i^\text{syn} \mid \beta &\sim \text{Poisson}(\mu_i) \nonumber \\
\text{with} \quad \text{log}(\mu_i) &= X_i\beta. \label{poismodel}
\end{align}
{Thus the mean of $f_i^\text{syn}$, denoted by $\mu_i$, is determined by $\beta$, the vector of log-linear model parameters ($X$ is the design matrix).} \par {The synthesizer must decide which interactions to include. This affects which relationships are preserved in the synthetic data. For example, if an all two-way interaction model is fitted, then relationships between all pairs of variables would be preserved, but higher order interactions - more complex relationships - would be lost. At the extreme, the saturated log-linear model includes all interactions and, as a result, each cell count in the multi-way table has its own parameter; see \cite{Agresti2013}. While the saturated model has} little value when {used for inference or prediction}, {including all} interaction{s} {does ensure that all associations are preserved in the synthetic data.} \par The model's minimal sufficient statistics are the observed marginal tables for the highest order terms included in the model{, f}or example, in the all two-way interaction model, all the observed two-way marginal tables. Practically, this means that the synthesizer does not require access to the full original table when synthesizing the data in this way. \par As {mentioned earlier}, expected counts from the fitted model can be obtained {relatively} quickly via the IPF algorithm. {The \code{syn.ipf} function in the \proglang{R} package \pkg{synthpop} \citep{Nowok2016} implements IPF, allowing the user this choice of synthesis method.} \subsubsection{Hierarchical Poisson log-linear model} \label{hierpois}
{A} hierarchical Poisson log-linear model {can also be used} to synthesize {the counts in the multi-way table, as proposed by \cite{Graham2007}.} The $i$th synthetic cell count $(i=1,\hdots,K)$ of the multi-way table $f_i^\text{syn}$, is modelled as follows:
\begin{align}
f_i^\text{syn} \mid \lambda_i &\sim \text{Poisson}(\lambda_i) \nonumber \\
\lambda_i \mid \beta, \xi &\sim \text{Gamma}(\xi, \xi/\mu_i), \ \mbox{with} \ \text{log}(\mu_i) = X_i\beta.
\end{align}
{The mean of $f_i^\text{syn}$, denoted by $\lambda_i$, is now assumed to be Gamma distributed; the mean of which, in turn, is determined by $\beta$, the vector of log-linear model parameters. The parameter $\xi$ affects the variance.} The marginal distribution of $f_i^\text{syn}$, { found by integrating over $\lambda_i$,} is the negative binomial distribution. \par {There are several sources of uncertainty in this model,} including model uncertainty, as decisions are made as to which interactions to include in $\beta$. {Moreover, \cite{Graham2007} generate fully synthetic data using the Bayesian posterior predictive distribution, so the $\beta$ are also assumed to be stochastic.}
\subsection{Challenges faced when synthesizing administrative data}
{The statistical challenges of dealing with administrative data are well documented; see \cite{Hand2018}.} And there are {further} challenges when synthesizing administrative data, in addition to the usual challenges faced when synthesizing {any} data {set}, such as finding the optimal balance between risk and utility. \par  Henceforth it is assumed that all variables in {an} administrative database are categorical. This assumption is not as strong as it might first appear. Continuous variables are often subject to rounding, for example, ages given as integers; {or} they can be categorized by the synthesizer, {for example,} ages can be \textit{converted} to integers. Besides, in any data set ($n$ {individuals}), the continuous variables take a finite number of values (maximum of $n$ values). This assumption allows the data to be expressed as a multi-way table.
\subsubsection{Large data sets} 
{In microdata format, administrative databases} are typically much larger than traditional survey data sets, {especially} in terms of the number of records {$n$} {(the number of rows in the microdata)}. \par {For categorical data sets expressed as a multi-way table, the data's size is not governed by rows and columns - but by cells. Neither the number of individuals $n$, nor the number of variables $p$, affects the table's dimensions. Instead, the dimensions are determined by the number of categories across the variables. As such, when synthesizing the data at the aggregated level, large $n$ can be beneficial, because it likely reduces the number of cells with zero counts (zero cells), and therefore relieves some of the problems caused by zero cells, discussed in Section \ref{addsmooth}.} \par {Administrative databases can include categorical variables with many categories. When fitting models such as the Poisson log-linear model, this} increases the number of parameters to estimate, {thus} causing the computational time to increase. It may be infeasible to fit models in such cases. The issue with computational time also extends to post-synthesis evaluations that are essential in examining the synthetic data's risk and utility. For example, the Bayesian estimation of disclosure risk given in \cite{Reiter2014} is computationally intensive even for relatively small survey data sets, as it involves continually re-fitting the synthesis model for every individual in the data, {and} ways such as importance sampling have already been incorporated to save time.
\subsubsection{Random and structural zeros} 
The presence of large categorical variables inevitably means that multi-way tables are sparse, that is, {they} have a high proportion of cells with zero counts. The zero cells can be said to consist {of} two sorts, {\textit{random zeros} and \textit{structural zeros}}. {Random zeros} are zero cells that arise through random chance: an individual with a given set of characteristics could have occurred but did not in the observed data. Structural zeros (as discussed in \citealt{Bishop1975}) are zeros that arise because a given set of characteristics is not possible, for example, a child aged three attending a secondary school. {When modelling contingency tables, structural zeros are usually dealt with by either removing the offending rows from the data set (there is no need for a balanced design when analysing contingency tables), or by weighting them out by incorporating a weight variable with weights of zero. } \par It is desirable that post-synthesis, all structural zeros remain zero, and some random zeros are transformed into non-zero counts. {Whenever random zeros are never synthesized to non-zeros, but some non-zeros are synthesized to zeros, it results in an inflated number of zeros in the synthetic data. This issue is discussed in greater detail in Section \ref{addsmooth}.} \par {In many models, such as in a Poisson log-linear model that is not saturated, it is difficult to account for structural zeros because cell means are smoothed to become non-zero. There are some exceptions, including the synthesis mechanism described by \cite{Manrique-Vallier2018}, which extends the non-parametric latent class model to account for structural zeros.}
\subsubsection{{No} defined sampling frame} \label{nosamplingframe}
Typically, administrative data are more akin to a census than a survey, and {thus} more akin to population data than sample data. {However, the population from which the data are drawn is unlikely to be well-defined, nor are the data likely to} constitute a simple random sample {from this unclear} population. For example, the  English School Census{, an administrative database held by the Department for Education,} includes pupils who attend state schools, but those who attend privately funded schools are excluded. \par {In general, obtaining inferences from administrative data requires careful consideration, and there are further considerations for synthesis. It can restrict the type of synthetic data which can be produced, as it may not be possible to generate fully synthetic data in the sense of \cite{Raghunathan2003}, which requires the generation of a synthetic population. This is difficult when the population in question is not obvious.}  \par {Moreover, risk evaluations in SDC often revolve around estimating {the probability that an} individual {who is} unique in the sample is also unique in the population \citep{Skinner1994}.} The{se} well established notions of \textit{sample uniqueness} and \textit{population uniqueness} become hazy when dealing with administrative data. 
\section{{The $(\sigma, \alpha)$-synthesis mechanism}} \label{section3}
Modelling for the purpose of generating synthetic data holds a unique position within statistical modelling: the objective is neither inferential nor predictive. Instead, the objective is solely to obtain synthetic data that resemble the original, but where disclosure risks are sufficiently low. That is, the model itself is not of interest. The use of saturated models, as proposed here, exploits this notion, alongside the notion that synthetic data {can be} obtained by sufficiently diverging away from the original data. {The original data itself can be viewed as having maximum utility - but also maximum risk. A synthetic data set can be generated by trading utility for disclosure protection, so that an acceptable balance between risk and utility is achieved.}   \par
Using saturated models helps to avoid the loss of relationships between variables. {In multiple imputation, w}hen the imputation model is less complex than the analyst's model{ subsequently fitted to the data}, the analyst's model is said to be ``uncongenial'' to the imputation model \citep{Meng1994}. {The same applies to synthesis models.} Therefore, over-fitting is preferable to under-fitting, and fitting saturated models is over-fitting in its most extreme.  \par Moreover, saturated synthesis models eliminate bias and model uncertainty. {Firstly, this means that the synthetic counts have an unbiasedness property: the expected counts in the synthetic data are equal to those in the original data}. {Secondly, it means that there are} fewer sources of uncertainty{: there is only one source - that from simulation}{. These two points} mean that the synthesizer has greater control of the {synthesis mechanism, because} certain properties of the synthetic data can be {derived} analytically, rather than needing to be found empirically. As synthetic data generation is an iterative process - data sets are generated, evaluated, improved upon and then regenerated - these analytical properties can improve the efficiency of the synthesis. \par {The \code{syn.catall} function in the \proglang{R} package \pkg{synthpop} \citep{Nowok2016} facilitates the use of a saturated multinomial model to produce synthetic data. Here, count models are considered instead.}
\subsection{{Synthesis through saturated count models (introducing $\sigma$)}}
\subsubsection{The Poisson model}
Suppose a saturated Poisson log-linear model is fitted to the {original data's} entire {multi-way} table. {Then each count in the multi-way table is assumed to be independent and Poisson distributed, with mean equal to the observed count.} Synthetic counts can be generated by simulating from these Poisson random variables, which adds stochastic error to the original counts and masks their true values. \par {The $i$th synthetic cell count $f_i^\text{syn}$ $(i=1,\hdots,K)$ of the multi-way table, is modelled as follows:}
\begin{align}
f_i^\text{syn}\mid \mu_i &\sim \text{Poisson}(\mu_i) \nonumber \\
\text{with} \quad \mu_i &= f_i, 
\end{align}
where $f_i$ is the corresponding count in the original data $(i=1,\hdots,K)$. {The difference with this model, compared to the model in (\ref{poismodel}), is that the $i$th synthetic count's mean $\mu_i$ is just equal to the original count $f_i$.} \par {Properties of the synthesis mechanism relate directly to properties of the Poisson distribution. For example, t}he Poisson distribution's probability mass function {gives} the probability that an arbitrary synthetic count $f^\text{syn}$ equals $N_2$, given that the original count $f$ equals $N_1$:
\begin{align*}
p(f^\text{syn}=N_2 \mid f=N_1) &=\frac{\text{exp}({-N_1})N_1^{N_2}}{{N}_2!}, 
\end{align*}
where $N_1$ and $N_2$ are non-negative integers. \par While the Poisson distribution is degenerate when the mean is zero, practically, this does not {affect} the method: whenever {an} original count is zero, the synthetic count is also zero. {Conveniently, t}his feature naturally accounts for structural zeros, which {rightly} remain zero.{ Random zeros, on the other hand, do need to be accounted for; a proposed solution is given in Section \ref{addsmooth}.} \par {This synthesis mechanism produces }completely synthesized {data using} the terminology of \cite{Raab2016}. However, somewhat confusingly as {a synthetic population is not created, the{ data are} partially synthetic in the sense of \cite{Reiter2003}, rather than fully synthetic in the sense of \cite{Raghunathan2003}; incidentally, \cite{Drechsler2018} seeks to clear up some of the confusion surrounding the term ``fully synthetic'' data sets.} Finally, the synthesis is via the ``plug-in approach'' \citep{Reiter2012}, that is, the Bayesian posterior predictive distribution is not used: synthetic {counts} are {simulated} {directly} from the {fitted} model. \par {It follows, naturally, that the expected ``sample'' size of the synthetic data $n_\text{syn}$ is equal to $n$, the sample size of the original data. The value $n_\text{syn}${, which is stochastic,} is the sum of the cell counts in the multi-way table - the table's grand total; and, as these counts are independent Poisson random variables whose means sum to $n$, $n_\text{syn}$ is also a Poisson random variable with mean $n$. Yet it need not be the case that $\mathbb{E}[n_\text{syn}]=n$. As \cite{Raab2016} demonstrate, {in completely synthesized data, }$n_\text{syn}$ can be made higher or lower than $n$. {Rather than the Poisson, the multinomial can be fit here using the same framework, which would guarantee that $n_\text{syn}=n$. Although it is worth considering whether fixing $n_\text{syn}$ is necessary - or even appropriate - with synthetic administrative data. Unlike a census, which has a known population total, an administrative database is unlikely to derive from a well-defined population, as discussed earlier in Section \ref{nosamplingframe}.} 
\subsubsection{Two-parameter count distributions for synthesis}
{T}he Poisson variability (variance equal to the mean), may not provide sufficient protection to at-risk records in the original data. {T}he variability can be increased - without introducing bias - {by} using overdispersed count {distributions} in place of the Poisson. {In a modelling context, these distributions, such as the negative binomial (NBI),} are suitable when{ever} the sampling variance exceeds that which is expected from the Poisson. \par 
{The hierarchical Poisson log-linear model given in Section \ref{hierpois} essentially assumes the cell counts follow a NBI distribution. A saturated model can be fit here, too, by again including all interaction effects. The shape parameter (denoted by $\sigma$ below) is set by the synthesizer, not least because there is insufficient degrees of freedom to estimate this parameter through maximum likelihood estimation. The notion is that the synthesizer determines, \textit{a priori} - that is, prior to synthesis - the variability required to achieve a pre-specified desired level of privacy and then adjusts the variance accordingly.} \par {When the NBI model is used, the $i$th synthetic cell count $f^\text{syn}_i$ $(i=1,\hdots,K)$ of the multi-way table, is modelled as follows:}
\begin{align}
f^\text{syn}_i \mid \mu_i, \sigma &\sim \text{NBI}(\mu_i, \sigma) \nonumber \\
\text{with} \quad \mu_i &= f_i. \nonumber
\intertext{
As with the Poisson, the NBI's probability mass function gives the probability that a{n arbitrary} synthetic count $f^\text{syn}$ equals $N_2$, given the original count $f$ equals $N_1$,
}
p(f^\text{syn}=N_2 \mid f=N_1, \sigma) &=  \frac{\Gamma(N_2+{1}/{\sigma})}{\Gamma(N_2+1) \cdot \Gamma({1}/{\sigma})} \cdot \bigg(\frac{\sigma N_1}{1+\sigma N_1}\bigg)^{N_2} \cdot \bigg(\frac{1}{1+\sigma N_1}\bigg)^{1/\sigma}.  \label{rose}
\intertext{The mean and variance of $f^\text{syn}$ are given as:}
\mathbb{E}[f^\text{syn}  \mid f=N_1, \sigma]&=N_1,\ \mbox{and} \  \text{Var}[f^\text{syn}  \mid f=N_1, \sigma]=N_1 + {\sigma}{N_1^2}, \label{meanvar}
\end{align}
which shows how the parameter $\sigma$ controls the variance of the model. {There is an array of two-parameter count distributions that can be used here. The other that is considered in this paper is t}he Poisson-inverse Gaussian (PIG) distribution (see \citealt{Rigby2019}):
\begin{align}
f^\text{syn}_i \mid \mu_i, \sigma &\sim \text{PIG}(\mu_i, \sigma) \nonumber \\
\text{with} \quad \mu_i &= f_i, \nonumber
\intertext{and, again, the probability that an arbitrary synthetic count $f^\text{syn}$ equals $N_2$, given that the original count $f$ equals $N_1$ is:}
p(f^\text{syn}= N_2 \mid f=N_1,\sigma)&=\bigg(\frac{2c}{\pi}\bigg)^{1/2} \cdot \frac{ N_1^{N_2} \text{exp}({1/\sigma}) K_{N_2-1/2}(c)}{(c \sigma)^{N_2} N_2!}, \label{rose1} \\
\text{where} \quad c^2=\frac{1}{\sigma^2}+\frac{2 N_1}{\sigma} \quad
&\text{and} \quad K_{\lambda}(t)=\frac{1}{2}\int\limits_0^{\infty} x^{\lambda-1} \exp\bigg\{-\frac{1}{2}t(x+x^{-1})\bigg\} \;\text{d}x \nonumber
\end{align}
is the modified Bessel function of the third kind. \par {The} parameterisations are as presented in {the} \proglang{R} package \pkg{gamlss.dist} \citep{Stasinopoulos2007}. The NBI and PIG distributions are both continuous mixtures of Poisson distributions. In the NBI the mixing distribution is the Gamma; in the PIG the mixing distribution is the inverse Gaussian distribution. They have identical mean and variance functions {(given in \ref{meanvar}), however higher moments differ.} The key point from these two-parameter distributions is that there is a parameter, $\sigma$, that is set by the synthesizer.
\subsection{Dealing with zero counts through additive smoothing (introducing $\alpha$)}  \label{addsmooth}
There is a downside with using saturated models that needs addressing: as there is no smoothing, zero cells in the original data {are always synthesized to zero}, result{ing} in too many zeros {in the synthetic data}. {That is, there are the zeros from the original data}, plus some non-zero cells that become zero through simulation. {An} excess of zero cells {can} affect the {risk and} utility of the synthetic data. \par
{With regards to risk, t}he issue is not so much with the zero cells themselves, which are relatively low risk, but {with what can be deduced from the non-zero} cells. {It follows that any non-zero cell in the synthetic data must have originated from a non-zero cell. So, from a non-zero cell in the synthetic data, an attacker can ascertain that \textit{at least one} individual belonged to that same cell in the original data.} \par
The addition of a pseudocount $\alpha>0$ {(which despite its name is not typically an integer)} to all {random} zeros in the original data (structural zeros should remain zero) opens the possibility that zero counts are synthesized to non-zeros. For example, when {the Poisson model is used and} $\alpha>0$ is added, the probability that a random zero $f=0$ is synthesized to $f^\text{syn}=N_2$ is:
\begin{align}
p(f^\text{syn}=N_2 \mid f=0, \; \alpha)&=\frac{\text{exp}({-\alpha})\cdot \alpha^{N_2}}{N_2!}. \nonumber 
\end{align} 
{The \code{syn.catall} function in \pkg{synthpop} \citep{Nowok2016}, which, as mentioned earlier, uses a saturated multinomial to produce synthetic data, allows the synthesizer to specify a Dirichlet prior; this is analogous to the addition of a pseudocount proposed here.}
\subsection{Tuning $\sigma$ and $\alpha$ to satisfy metrics \textit{a priori}}
{The upshot of this synthesis mechanism is that there are two parameters, $\sigma$ and $\alpha$, that are controlled by the synthesizer. These can be tuned to satisfy certain risk or utility metrics. The following $\tau$ metrics, are an example of a simple set of metrics that can not only be tuned by $\sigma$, but can also be represented analytically.} 
\begin{enumerate}
\item $\tau_1(k):=$  The proportion of cells of size $k$ in the synthetic data.
\item $\tau_2(k):=$ The proportion of cells of size $k$ in the original data.
\item $\tau_3(k):=$ The proportion of cells of size $k$ in the original data, which remain of size $k$ in the synthetic data.
\item $\tau_4(k):=$ The proportion of cells of size $k$ in the synthetic data, which were also of size $k$ in the original data.
\end{enumerate}
Metrics $\tau_1$, $\tau_2$ and $\tau_4$ are conditional on the distribution of cell sizes in the original data{'s multi-way table}, whereas $\tau_3$ is not. To illustrate, suppose {a} data set is comprised entirely of cell counts of one and that the Poisson distribution is used to generate synthetic counts{. T}hen $\tau_3(1)=\text{exp}({-1})${, which is given by the Poisson's probability mass function;} and $\tau_4(1)=1${, because the original data contains only ones}. {Now, when} an original data set comprises a range of non-zero counts, then $\tau_3(1)=\text{exp}({-1})$ remains unchanged, but $\tau_4(1)\leq1$ because a synthetic cell count of one could have originated from any non-zero cell. \par {As structural zeros are not involved in the synthesis and just remain zero, when $k=0$ these $\tau$ metrics refer to random zeros, for example, $\tau_1(0)$ is the proportion of random zeros in the synthetic data.} \par The expected values of these $\tau$ metrics can be derived analytically, as demonstrated below for when the Poisson model is used for synthesis.
\subsubsection{The metrics $\tau_1$ and $\tau_2$}
The metric $\tau_1(k)$ is the proportion of cells of size $k$ in the synthetic data, that is,
\begin{align}
\tau_1(k)&=p(f^{\text{syn}}=k) \quad k=0,1,2, \hdots \\
\intertext{which, by the law of total probability,} \nonumber
&=  \sum\limits_{j=0}^\infty p(f^{\text{syn}}=k \mid f=j) \cdot p(f=j) \ = \ \frac{\text{exp}({-\alpha})\alpha^k}{k!} \cdot \tau_2(0) + \sum\limits_{j=1}^\infty \frac{\text{exp}({-j})j^k}{k!} \cdot \tau_2(j), \nonumber  
\intertext{where $\tau_2(k)$ is simply the proportion of cells with a count of $k$ in the original data denoted by,}
\tau_2(k)&=p(f=k) \quad k=0,1,2, \hdots\;.
\end{align}
\subsubsection{The metric $\tau_3$}
The metric $\tau_3(k)$ gives the proportion of cells of size $k$ in the original data, which remain of size $k$ in the synthetic data, that is,
\begin{align}
\tau_3(k)&=p(f^{\text{syn}}=k | f=k) \   \quad k=0,1,2, \hdots\;  \\
&=  \begin{cases}
   {\text{exp}({-\alpha})\alpha^k}/{k!}  & \text{if $k=0$ } \\
     {\text{exp}({-k})k^k}/{k!} & \text{if $k\geq1$}
  \end{cases} \nonumber 
\intertext{
\subsubsection{The metric $\tau_4$}
The metric $\tau_4(k)$ is the proportion of cells of size $k$ in the synthetic data, which were also of size $k$ in the original data, that is, }
\tau_4(k)&=p(f=k | f^{\text{syn}}=k) \quad k=0,1,2, \hdots\; \intertext{The metric $\tau_4(k)$ can be expressed in terms of the other $\tau$ metrics:}
\tau_4(k)&=p(f=k | f^{\text{syn}}=k)= \frac{ p(f^{\text{syn}}=k | f=k) \cdot p(f=k)} {p(f^{\text{syn}}=k) } =\frac{\tau_3(k) \cdot \tau_2(k)}{\tau_1(k)} \nonumber \\
&=\begin{cases}
   {\text{exp}({-\alpha})\alpha^k} \cdot \tau_2(0)\bigg/\bigg({{\text{exp}({-\alpha})\alpha^k} \cdot \tau_2(0) + \sum\limits_{j=1}^\infty {\text{exp}({-j})j^k} \cdot \tau_2(j)}\bigg) & \text{if $k=0$ } \\
     {\text{exp}({-k})k^k} \cdot \tau_2(k)\bigg/\bigg({{\text{exp}({-\alpha})\alpha^k} \cdot \tau_2(0) + \sum\limits_{j=1}^\infty {\text{exp}({-j})j^k} \cdot \tau_2(j)}\bigg) & \text{if $k\geq1$}
  \end{cases} \nonumber 
\end{align}
\subsubsection{The $\tau$ metrics' link to disclosure risk}
The notion of disclosure risk is different for synthetic data in tabular format, than for microdata. When {microdata} are aggregated and synthesized, the direct links between individuals in the original and synthetic data are lost. \par 
Uniques are individuals who belong to a cell with a count of one, and are often considered to be most at risk of disclosure. An important value with respect to risk is $\tau_4(1)$: the proportion of uniques in the synthetic data which {we}re also unique in the original data. This is arguably more important than $\tau_3(1)$: the proportion of uniques in the original data which are also unique in the synthetic data. This is because the former assumes knowledge of the synthetic data, which an attacker has access to; {whereas t}he latter assumes knowledge of the original data, which an attacker cannot access. \par
{There are two ways in which the synthesizer can reduce $\tau_4(1)$. The first way is to increase $\sigma$, which increases the variance of the synthetic counts. The second way is to increase $\alpha$. Zero cells with {small }$\alpha>0$ added are much more likely to be synthesized to one than to any other non-zero value. For example, when $\alpha=0.1$ and {the Poisson model is used}, a zero count is exactly twenty times more likely to be synthesized to one than two, which increases the number of uniques in the synthetic data and thereby decreases $\tau_4(1)$.} 
\subsubsection{Tuning $\sigma$ and $\alpha$ to adjust the expected values of the $\tau$ metrics}
{The notion is that the synthesizer tunes $\sigma$ and $\alpha$ to yield synthetic data with certain properties. As an example, the synthesizer can decide, \textit{a priori}, that they would like $\tau_1(0)=\tau_2(0)$ and $\tau_4(1)=p$ (for some $p$) - and they can then tune $(\sigma, \alpha)$ accordingly.} 

\par  When $\alpha=0$, the inequality $\tau_1(0) \geq\tau_2(0)$ holds, since zero cells in the synthetic data comprise all zero cells in the original data, plus those that randomly become zero through synthesis. As $\alpha$ increases, the {expected} difference between $\tau_1(0)$ and $\tau_2(0)$ narrows and the inequality would eventually reverse. \par An attractive property might be for the synthetic data to have the same proportion of zero cells as the original data, that is, for $\tau_1(0)=\tau_2(0)$. Under the Poisson model this is achieved by setting:
\begin{align}
\alpha^* &= -\text{log}\bigg\{1- \frac{1}{\tau_2(0)}\sum\limits_{j=1}^\infty \text{exp}({-j}) \cdot \tau_2(j) \bigg\}. \nonumber
\end{align}
{The short derivation is given in the supplementary material.} An alternative is to choose $\alpha$ such that $\tau_4(1)=p$, where the synthesizer decides what $p \in [0,1]$ is acceptable (a pre-specified level of disclosure risk). 
Here the value of $\alpha^*$  must be obtained numerically; under the Poisson model it satisfies,
\begin{align}
p&= \text{exp}({-1})\cdot \tau_2(1) \bigg/\bigg({{{\alpha^*}\text{exp}({-\alpha^*})  }} \cdot  \tau_2(0) + \sum\limits_{j=1}^\infty {j\cdot\text{exp}({-j})}{} \cdot \tau_2(j) \bigg). \nonumber
\end{align}
{Similar expressions can be derived for the two-parameter synthesis models{, where the required value of $\alpha^*$ also depends on $\sigma$. These are also included in the supplementary material.}}
\section{Empirical study}
\subsection{The data}
{ The  English School Census (ESC) is an administrative database that holds information about pupils in state-funded schools. Every school term the Department for Education (DfE) requests that all nursery, primary and secondary schools, which are
fully or partly funded by the state, submit details about the school and its pupils. This is just one example of an administrative database held by a government department; other examples include, but are not limited to, the Patient Register (held by the Department of Health) and the Customer Information {System} (held by the Department for Work and Pensions).}

\par {For obvious reasons, access to the ESC data, as well as to other administrative databases, is highly restricted. However, in previous work conducted by the ONS, a carefully constructed data set using publicly available sources was created to be used as a substitute to the ESC, in order to develop synthesis methods for administrative data{. These data were used here as the basis for generating a synthetic database.} \par The data were generated using public 2011 census output tables involving various combinations of local authority, sex, age and ethnicity\footnote{Specifically, information from the following public sources were used to create the data: \\ {\tt http://www.nomisweb.co.uk/census/2011; http://www.ons.gov.uk/ons/guide-method/census/2011/
census-data/2011-census-user-guide/quality-and-methods/quality/quality-measures/ response-and-imputation-rates/index.html; https://www.gov.uk/government/statistics/ schools-pupils-and-their-characteristics-january-2014 }}. Language attributes from the census were also included and artificially expanded to match with categories in the ESC. In addition, school phase attributes were incorporated, some adjustments for migration were applied, and non-response and invalid categories were added to various variables, again taking publicly available information from the census.}    

The two variables measured at the school level were ignored for this illustration, which focused instead on the remaining five variables measured at the pupil level. Henceforth, this data set is referred to as the ESC{sub} where {``sub''} denotes {substitute}. Table \ref{datasummary} summarises the variables present in the ESC{sub} illustration. The data comprise $n=8,190,870$ pupils over $p=5$ categorical variables, giving rise to a multi-way contingency table with $K=326 \times 20 \times 4 \times 19 \times 7=3.5 \times 10^6$ cells. The breakdown of the cell counts are given in Table \ref{datasummary2}; only 333,660 (9.6$\%$) are non-zero - so the data are sparse. There are no structural zeros. \par So while the data are in a sense simulated, this was done using real data sources and care was taken to ensure that the resulting data reflect, at the very least, the typical structure present in the ESC. As such, this was a good example to use to demonstrate our synthesis method and a similar performance is expected when the method is applied to the actual ESC, as well as other similar large categorical administrative databases.  Importantly, the data were not generated from a statistical model and thus do not favour {a particular} synthesis method.  
    \begin{table}
\caption{\label{datasummary}The ESC{sub}'s variables and their numbers of categories.}
\centering
\fbox{
\begin{tabular}{*{3}{c}}
\em Variable & \em Type & \em \# Categories\\
\hline
Area Code/ Geography (V) & Categorical & 326\\
Ethnicity (W) & Categorical & 20  \\
Sex (X) & Categorical & 4 \\
Age (Y) & Categorical & 19 \\
Language (Z) & Categorical & 7  \\
\end{tabular}}
\end{table} 
\begin{table}
\caption{\label{datasummary2}Distribution of cell sizes in the ESC{sub} data.}
\centering
\fbox{
\begin{tabular}{*{3}{c}}
Cell count & Frequency & $\%$ of cells \\
\hline 
0 & 3,134,980 & 90.38 \\
1 & 119,917 & 3.46 \\ 
2 & 51,412 & 1.48 \\ 
3 & 25,952 & 0.75 \\
4 & 19,450 & 0.56 \\
5 & 13,076 & 0.38 \\
6 & 10,345 & 0.30  \\
7 & 7,947 & 0.23 \\
8 & 7,077 & 0.20   \\
9 &  5,809 & 0.17  \\
10 & 5,163 & 0.15 \\
$11 \leq$ & 67,512 & 1.95  \\  \hline
Total & 3,468,640 & 100
\end{tabular}}
\end{table}


\subsection{The synthesis}  
The synthesis was carried out in \proglang{R} (version 3.6.3) using the methods described in this paper. The Poisson, NBI and PIG models were compared {by examining how the counts in the synthetic data's multi-way table deviate from those in the original data, and by computing summaries of risk and utility. This evaluation also includes a comparison of parameter estimates obtained from a log-linear analysis, which was performed on both the original and the synthetic data.} \par {J}ust $m=1$ data set was generated {for each synthesis model}. The CPU times to carry this out were 0.2, 0.3 and 162 seconds for the Poisson, NBI and PIG models, respectively. {The PIG model took notably longer, although is still fast compared to other methods, such as the conditional approaches (Section \ref{conditional}) that synthesize the data at the microdata level}. \par Let $V$, $W$, $X$, $Y$ and $Z$ denote the five variables in the data, and let $f_{vwxyx}$ denote the cell count of a particular cell in the cross-classified table corresponding to category $v \in V$, $w \in W$ , $x \in  X$, $y \in  Y$ and $z \in  Z$. A synthetic count was then drawn for this cell by,
\begin{align}
f^\text{syn}_{vwxyz} &\sim \text{Poisson}(f_{vwxyz}), \nonumber
\intertext{when the  Poisson synthesis model was used; or, when either of the two-parameter distributions were used,}
f^\text{syn}_{vwxyz} &\sim \text{NBI}(f_{vwxyz},\sigma) \quad \text{or} \quad f^\text{syn}_{vwxyz} \sim \text{PIG}(f_{vwxyz}, \sigma). \nonumber
\end{align}
For the Poisson, the only parameter to be set was $\alpha$, the pseudocount added to random zeros in the original data. For the two-parameter count models, there was the additional parameter $\sigma$ to consider. \par As mentioned previously, one of the appealing features of using these saturated synthesis models is that it allows the synthesizer to determine properties of the synthesis model \textit{a priori}, thus reducing the amount of empirical evaluation necessary during the synthesis. For illustration, Figure \ref{fig04} (left) compares the effect of $\sigma$ on the risk metric $\tau_4(1)$ for the NBI and PIG models{. F}or large $\sigma${,} the risk levels off for the NBI but continues to fall away for the PIG. Figure \ref{fig04} (right) also looks at $\tau_4(1)$, but at the combined effect of $\sigma$ and $\alpha$ when the NBI {is used. F}or all $\sigma$, $\tau_4(1)$ {falls} as $\alpha$ increases{ and $\tau_4(1)$ is always lower for the NBI than for the Poisson}.
\begin{figure}
\centering
\includegraphics{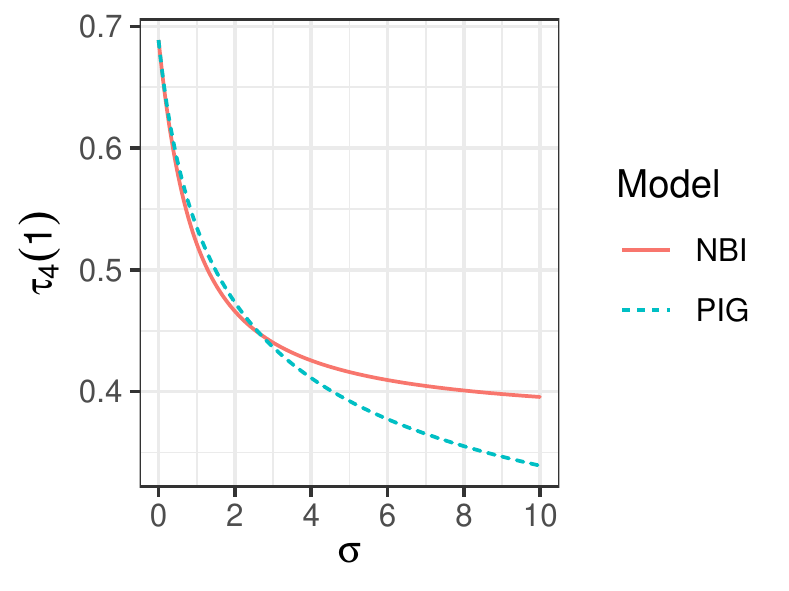}
\includegraphics{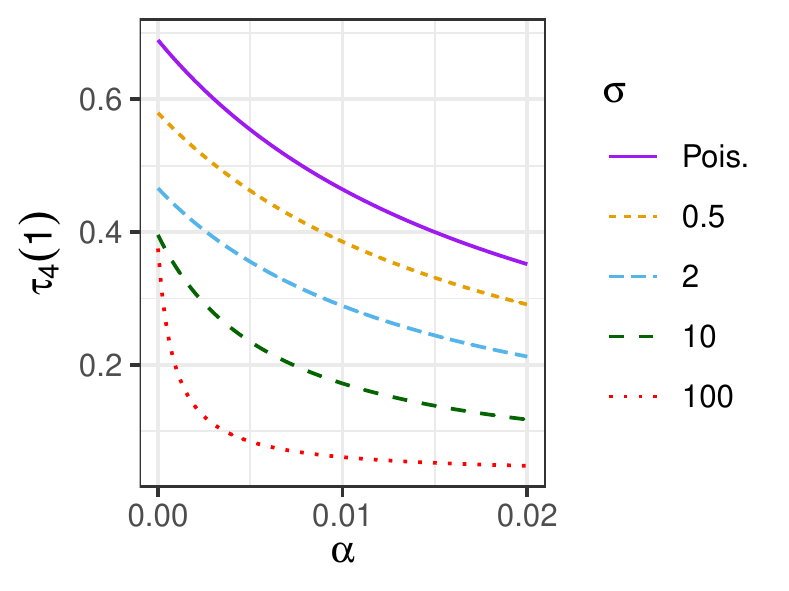}
\caption{\label{fig04} The left plot gives the risk metric $\tau_4(1)$ - the proportion of uniques in the synthetic data that were also unique in the original data - as a function of $\sigma$ for when the NBI (solid line) and PIG (dashed line) models were used ($\alpha=0$). The right plot shows how $\alpha$ and $\sigma$ together affect $\tau_4(1)$ when the NBI is used.}
\end{figure} 


\subsection{Descriptive summaries of risk and utility}

{Table \ref{tab04} gives the proportion of cell counts {in the synthesized tables} that are with{in} $p\%$ of their original size, for different $\sigma$ and $\alpha$. The first block of results considers all cells, while the second block only considers non-zero cells (in the observed data). Smaller values of $p$ can be viewed as summaries of risk while larger values of $p$ measures of utility. {To elaborate, if a large proportion of original and synthetic cell counts are very close, say within $0.5\%\;(p=0.5)$ of each other, then the synthetic data could be considered to be high risk. If, on the other hand, few original and synthetic cell counts are within, for example, $50\%\;(p=50)$ of each other, then this is likely to indicate low utility. As an example, the 0.927 value in the top-left corner of the table means that when $\alpha=0$ and $\sigma=0$, 92.7$\%$ of all cell counts in the synthetic data were within 0.5$\%$ of the corresponding count in the original data.}  }The Poisson model had the greatest utility but the greatest risk. There was little to choose between the NBI and PIG models based on these summaries. As expected, greater $\sigma$ or $\alpha$ lead to greater divergences in original and synthetic cell sizes. \par The {similarities between the NBI and PIG models} are also highlighted in Figures \ref{fig37} and \ref{fig47}, which plot the synthetic versus original counts {and also} percentage differences (between synthetic and original counts) versus original count{s}. While the Poisson model{'s} points ($\sigma = 0$ case{s}) were close to the $45^{\circ}$ line - {which indicates} strong correlation between synthetic and original counts - this correlation reduces as $\sigma$ increases in both the NBI and PIG models. {Even a relatively small value of $\sigma = 0.01$ introduced noticeable dispersion around the $45^{\circ}$ line.} The right panels display a funnel shape{, that is, percentage differences were greater for smaller counts than for larger counts.} This is an ideal profile for balancing risk and utility, as the riskiest individuals are the ones corresponding to small cell counts, {and these cell counts require the} most movement during synthesis. {On the other hand, large counts are relatively low risk, and proportional changes to large counts will have a more significant impact on utility, thus relatively less perturbation is desired.}

Table \ref{tab05} presents empirical values for the $\tau$ metrics, again for varying $\sigma$ and $\alpha$. The {expected} values are known prior to synthesis, though a small difference occurs, owing to simulation noise. But, for cell sizes that are prevalent in the original data - such as zeros and ones - this error is negligible. For example, the empirical value obtained for $\tau_3(1)$ when the Poisson model was used ($\sigma=0,\alpha=0$) is 0.3674, which is almost identical to the {expected} value, exp(-1)=0.3679. \par 
Table \ref{tab05} {also} illustrates the suitability of $\alpha$ in reducing risk. The values for $\tau_4(1)$ are substantially lower when $\alpha=0.02$ than when $\alpha=0$; for example, when the Poisson model is used, $\tau_4(1)$ is 0.352 compared to 0.689. For a given $\alpha$, the NBI and PIG models almost always have a lower risk than the Poisson model when considering the $\tau_3(1)$ and $\tau_4(1)$ metrics. It is particularly interesting to note varying profiles between synthesis models and these metrics. For example, if one model has a lower $\tau_3(1)$ value than another, then this is not necessarily the case when comparing the corresponding $\tau_4(1)$ value. To illustrate, consider the case when $\alpha = 0$ and $\sigma = 10$. For the NBI, the value of $\tau_3(1)$ is $0.0711 < 0.1532$ the value for the PIG. But for $\tau_4(1)$, with these same parameter values, the value under the NBI is $0.3910 > 0.3387$ the value under the PIG. The specific choice of synthesis model to use would depend on the synthesizer's range of permitted values for $\tau_3$, and $\tau_4$, and choosing the model that best satisfies these requirements.


\begin{center}
\begin{table}
\caption{\label{tab04} Empirical results showing the proportion of synthetic cell counts within $p\%$ of the corresponding original counts. The table includes results for both the NBI and the PIG, for different $\sigma$ and $\alpha$. The upper block of results considers all original cell counts, while the lower block considers only non-zero original cells. For $\alpha=0.02$, whenever a zero count was synthesized to a non-zero count, although the percentage difference was not estimable (zero denominator), it was deemed to be greater than $50\%$ for the purpose of this table.}
\small
\centering
\fbox{
\begin{tabular}{*{13}{c}}
&& \multicolumn{11}{c}{Proportion of synthetic cell counts within $p\%$ of the original} \\
&& \multicolumn{5}{c}{NBI} &  & \multicolumn{5}{c}{PIG}  \\
\cline{3-7}
\cline{9-13}
& $p$ & 0.5 & 1 & 5 & 10 & 50 &  & 0.5 & 1 & 5 & 10 & 50  \\ 
 \cline{3-7}
\cline{9-13} \\
& & \multicolumn{11}{c}{All original cell counts} \\
& $\sigma$\;\;\;\;\;  & \multicolumn{11}{c}{} \\
\multicolumn{1}{c|}{}&0 (Pois.) & 0.927 & 0.927 & 0.931 & 0.935 & 0.967 & &0.927 & 0.927 & 0.931 & 0.935 & 0.967  \\ 
\multicolumn{1}{c|}{} & 0.1 & 0.924 & 0.924 & 0.926 & 0.928 & 0.961 && 0.925 & 0.925 & 0.926 & 0.928 & 0.961  \\ 
 \multicolumn{1}{c|}{} &0.5 & 0.920 & 0.920 & 0.920 & 0.922 & 0.946 & &0.921 & 0.921 & 0.921 & 0.923 & 0.949  \\ 
\multicolumn{1}{c|}{$\alpha=0$}  &1 & 0.917 & 0.917 & 0.917 & 0.918 & 0.937 & &0.918 & 0.918 & 0.919 & 0.920 & 0.942  \\ 
\multicolumn{1}{c|}{}  &2 & 0.914 & 0.914 & 0.914 & 0.914 & 0.928  & &0.916 & 0.916 & 0.916 & 0.917 & 0.935  \\ 
\multicolumn{1}{c|}{}  &5 & 0.910 & 0.910 & 0.910 & 0.910 & 0.918  && 0.913 & 0.913 & 0.913 & 0.914 & 0.927  \\ 
\multicolumn{1}{c|}{}  &10 & 0.907 & 0.907 & 0.907 & 0.908 & 0.912  & &0.911 & 0.911 & 0.911 & 0.912 & 0.921  \\ \\
\multicolumn{1}{c|}{}&0 (Pois.) & 0.909 & 0.910 & 0.913 & 0.917 & 0.949  & & 0.909 & 0.910 & 0.913 & 0.917 & 0.949  \\ 
\multicolumn{1}{c|}{} & 0.1 & 0.907 & 0.907 & 0.908 & 0.910 & 0.943  & & 0.907 & 0.907 & 0.908 & 0.910 & 0.943  \\ 
\multicolumn{1}{c|}{}  &0.5 & 0.902 & 0.902 & 0.903 & 0.904 & 0.928  && 0.903 & 0.903 & 0.903 & 0.905 & 0.931  \\ 
 \multicolumn{1}{c|}{$\alpha=0.02$} &1 & 0.899 & 0.899 & 0.900 & 0.901 & 0.920  && 0.901 & 0.901 & 0.901 & 0.902 & 0.924  \\ 
\multicolumn{1}{c|}{}  &2 & 0.896 & 0.896 & 0.896 & 0.897 & 0.911  && 0.899 & 0.899 & 0.899 & 0.900 & 0.918  \\ 
 \multicolumn{1}{c|}{} &5 & 0.893 & 0.893 & 0.893 & 0.893 & 0.901  && 0.896 & 0.896 & 0.896 & 0.897 & 0.909  \\ 
\multicolumn{1}{c|}{}  &10 & 0.891 & 0.891 & 0.891 & 0.891 & 0.896  && 0.895 & 0.895 & 0.895 & 0.895 & 0.905  \\ \\
& & \multicolumn{11}{c}{Non-zero original cell counts} \\
& $\sigma$\;\;\;\;\;  & \multicolumn{11}{c}{} \\
\multicolumn{1}{c|}{}& 0 (Pois.) & 0.242 & 0.245 & 0.280 & 0.327 & 0.658  && 0.242 & 0.245 & 0.280 & 0.327 & 0.658  \\ 
 \multicolumn{1}{c|}{} &0.1 & 0.214 & 0.215 & 0.226 & 0.252 & 0.592  && 0.217 & 0.218 & 0.229 & 0.256 & 0.598  \\ 
 \multicolumn{1}{c|}{} &0.5  & 0.167 & 0.167 & 0.173 & 0.187 & 0.437  & &0.177 & 0.177 & 0.182 & 0.197 & 0.468  \\ 
 \multicolumn{1}{c|}{$\alpha=0$} &1 & 0.136 & 0.136 & 0.140 & 0.150 & 0.347  && 0.153 & 0.153 & 0.157 & 0.167 & 0.395  \\ 
 \multicolumn{1}{c|}{} &2 & 0.102 & 0.102 & 0.105 & 0.111 & 0.253  && 0.128 & 0.128 & 0.131 & 0.138 & 0.324  \\ 
 \multicolumn{1}{c|}{} &5 & 0.059 & 0.059 & 0.061 & 0.064 & 0.145  && 0.097 & 0.097 & 0.099 & 0.104 & 0.238  \\ 
 \multicolumn{1}{c|}{} &10 & 0.037 & 0.037 & 0.038 & 0.040 & 0.089  && 0.076 & 0.076 & 0.077 & 0.081 & 0.183  \\ \\
\multicolumn{1}{c|}{}& 0 (Pois.)  & 0.242 & 0.245 & 0.279 & 0.326 & 0.657  && 0.242 & 0.245 & 0.279 & 0.326 & 0.657 \\ 
  \multicolumn{1}{c|}{}& 0.1  & 0.215 & 0.215 & 0.226 & 0.253 & 0.593  && 0.215 & 0.216 & 0.227 & 0.254 & 0.598  \\ 
  \multicolumn{1}{c|}{}& 0.5  & 0.167 & 0.167 & 0.172 & 0.186 & 0.437  && 0.175 & 0.176 & 0.181 & 0.196 & 0.468  \\ 
  \multicolumn{1}{c|}{$\alpha=0.02$}& 1  & 0.137 & 0.137 & 0.141 & 0.151 & 0.348  && 0.153 & 0.153 & 0.157 & 0.167 & 0.396  \\ 
  \multicolumn{1}{c|}{}& 2  & 0.102 & 0.102 & 0.105 & 0.111 & 0.253  && 0.128 & 0.128 & 0.130 & 0.138 & 0.324  \\ 
  \multicolumn{1}{c|}{}& 5  & 0.061 & 0.061 & 0.062 & 0.065 & 0.147  && 0.096 & 0.096 & 0.098 & 0.103 & 0.237  \\ 
  \multicolumn{1}{c|}{}& 10  & 0.037 & 0.037 & 0.037 & 0.039 & 0.088  && 0.076 & 0.076 & 0.077 & 0.081 & 0.182  \\  \\
   \hline
\end{tabular}
}
\end{table}
\end{center}




\begin{figure}
\centering
\includegraphics{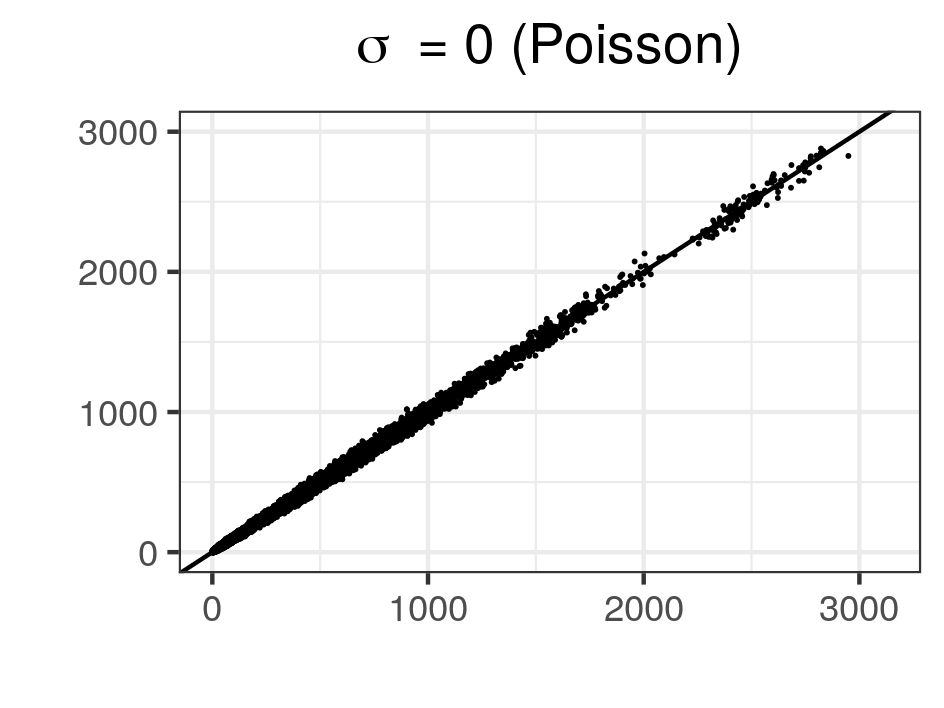}
\includegraphics{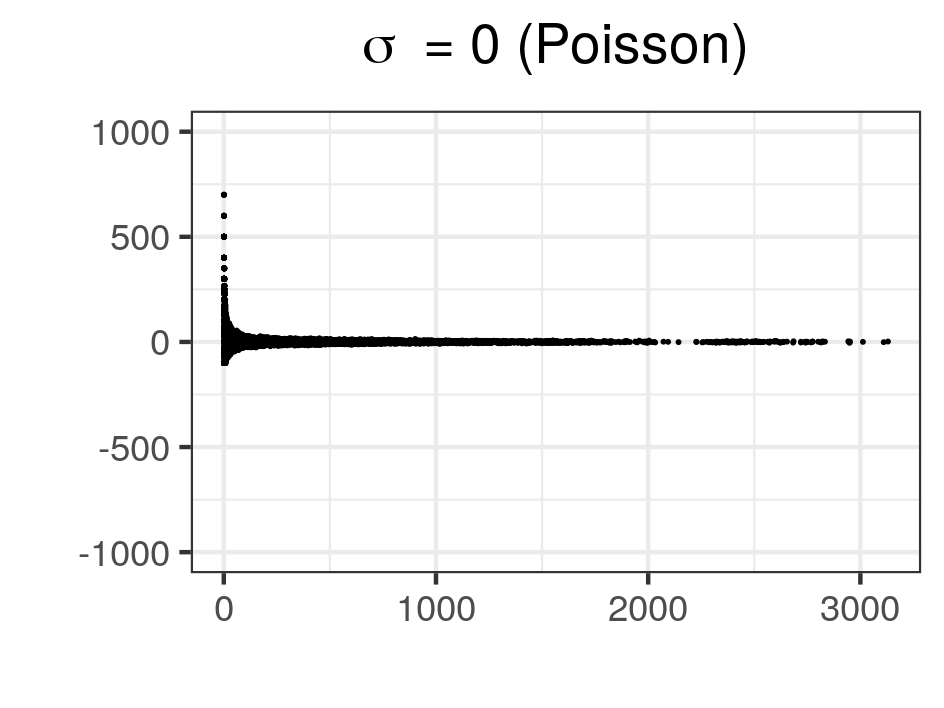}
\includegraphics{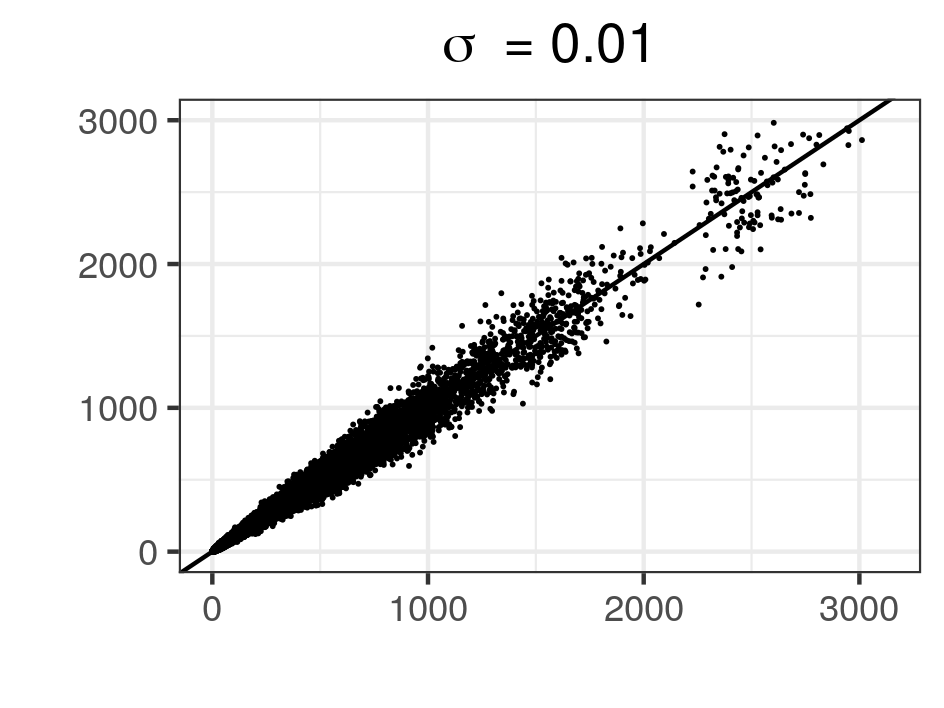}
\includegraphics{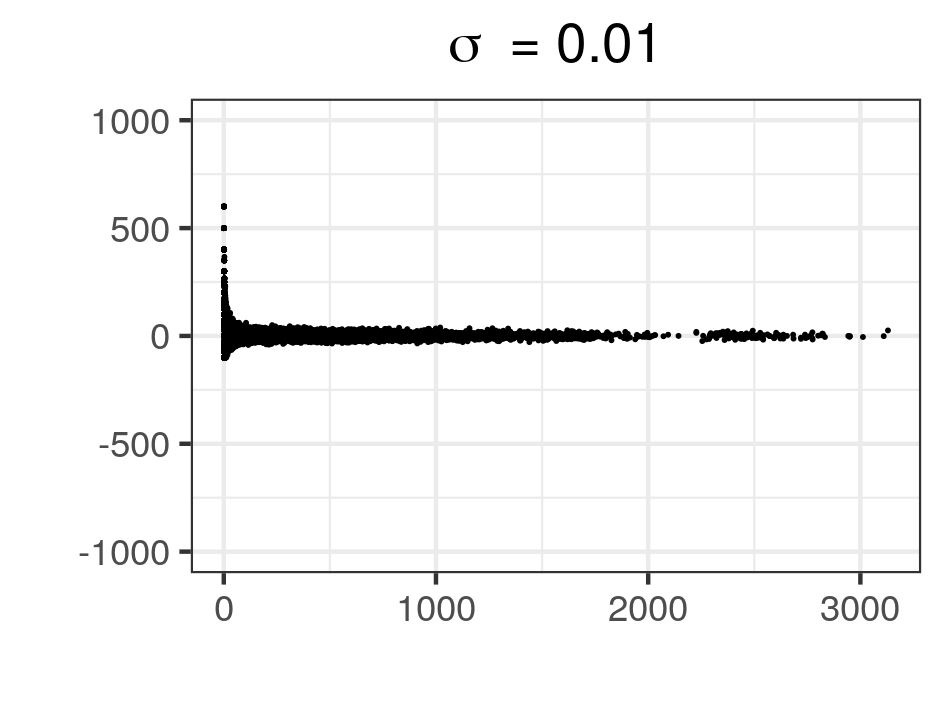}
\includegraphics{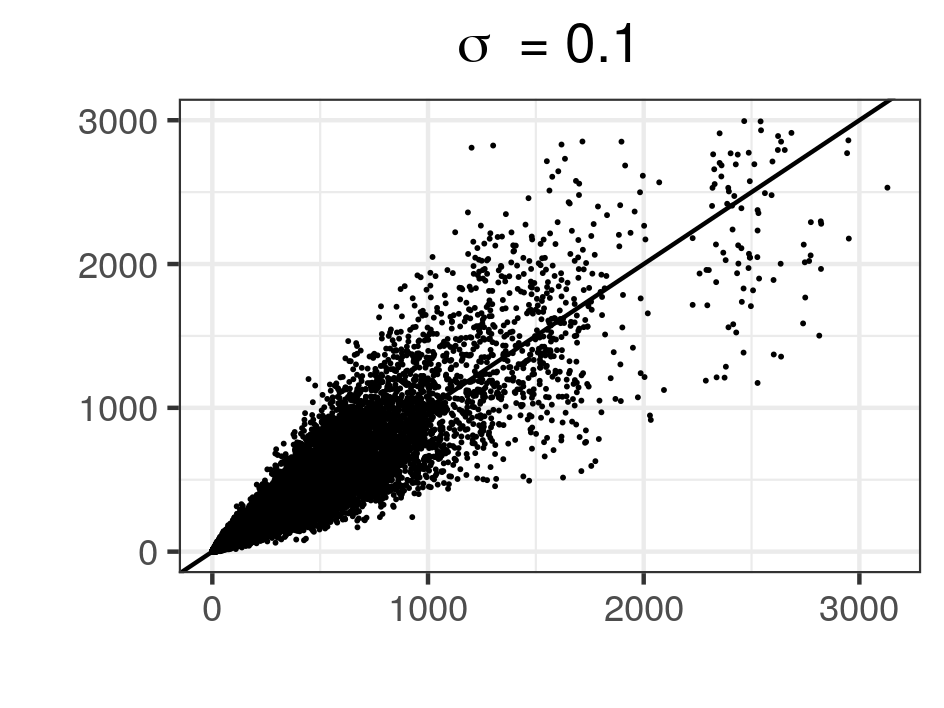}
\includegraphics{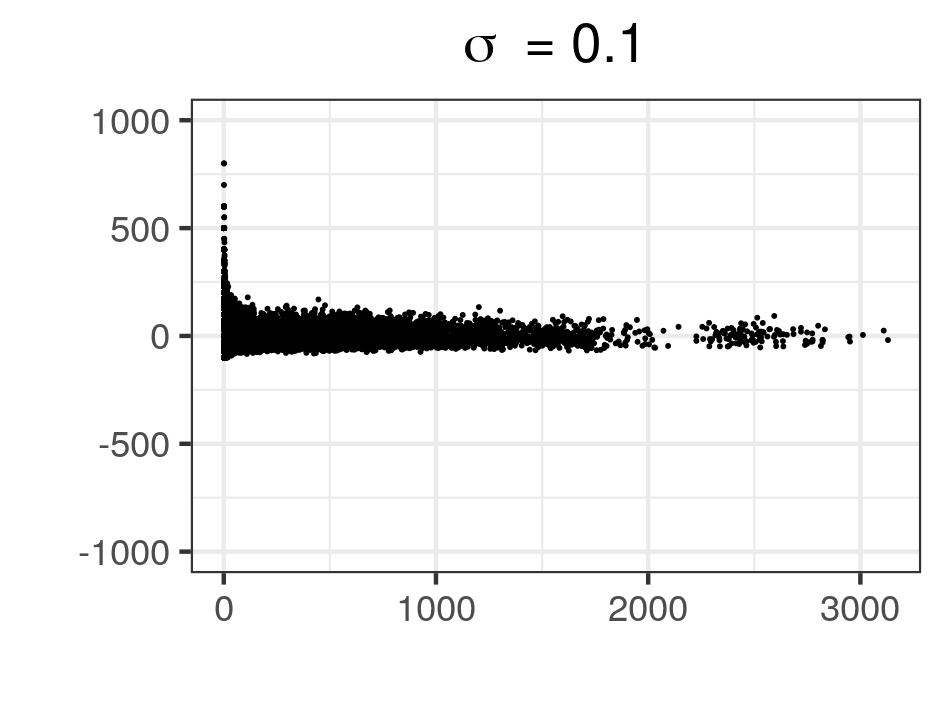}
\includegraphics{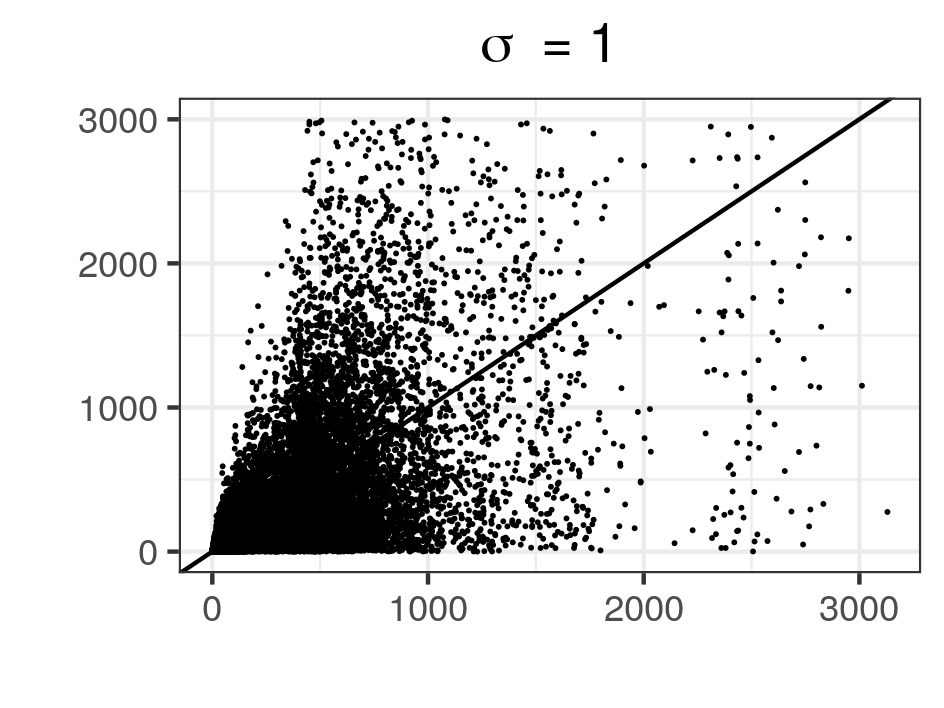}
\includegraphics{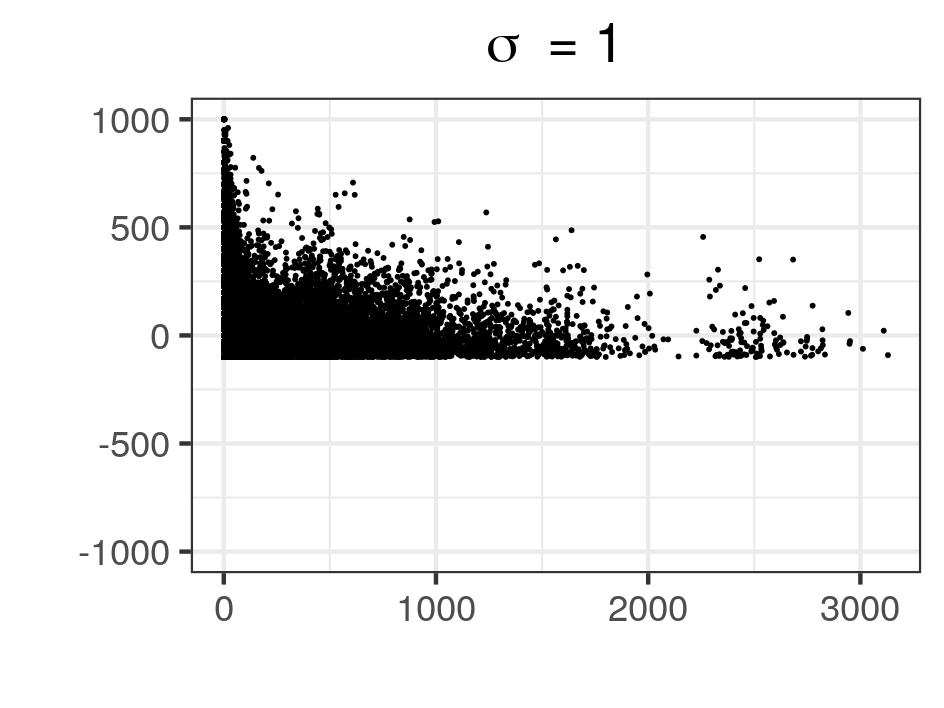}
\caption{\label{fig37} The left hand plots give the synthetic counts versus the original counts for different $\sigma$ when the NBI model wass used for synthesis{ and $\alpha=0$. The original counts of zero - which were always synthesized to zero because $\alpha=0$ - are omitted.} The right hand plots give original and synthetic counts' percentage differences versus the original counts. {The percentage differences were calculated by: $100*$(synthetic count - original count) / original count.}}
\end{figure}

\begin{figure}
\centering
\includegraphics{plot30.png}
\includegraphics{plot31.png}
\includegraphics{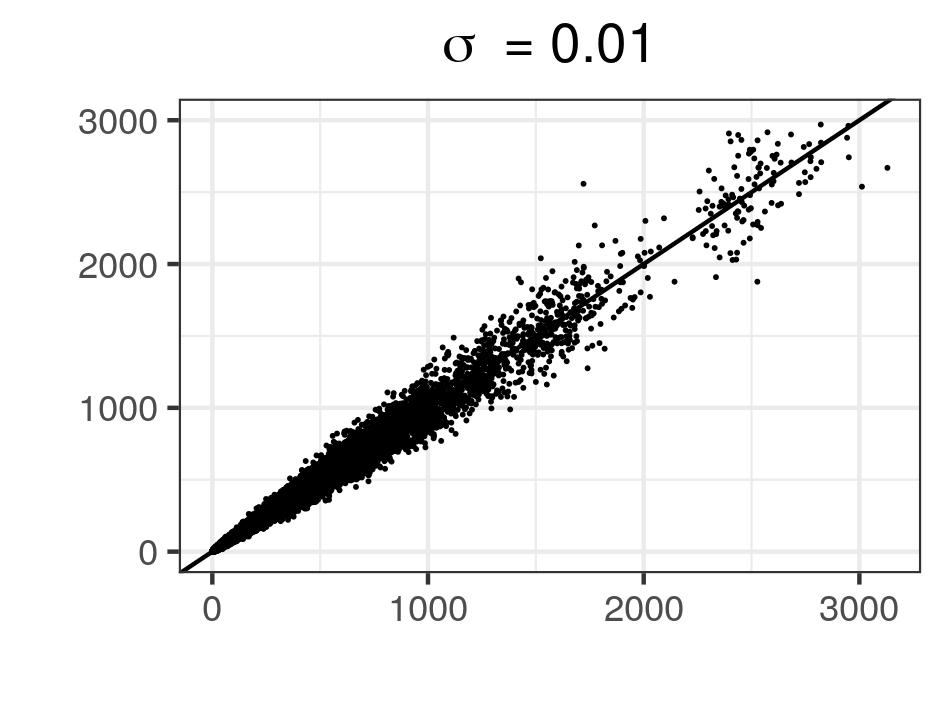}
\includegraphics{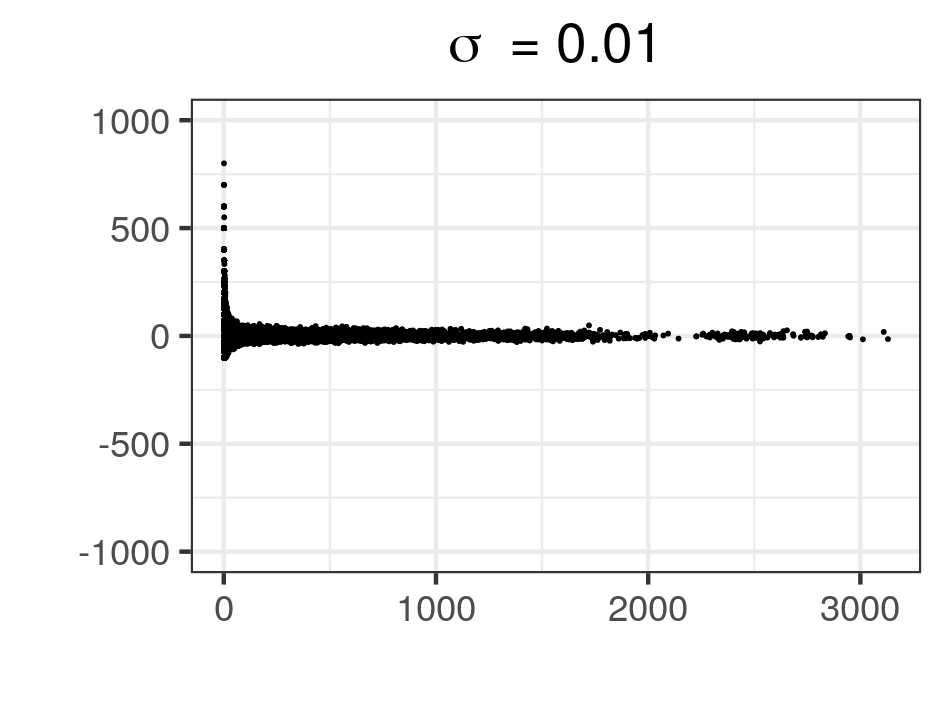}
\includegraphics{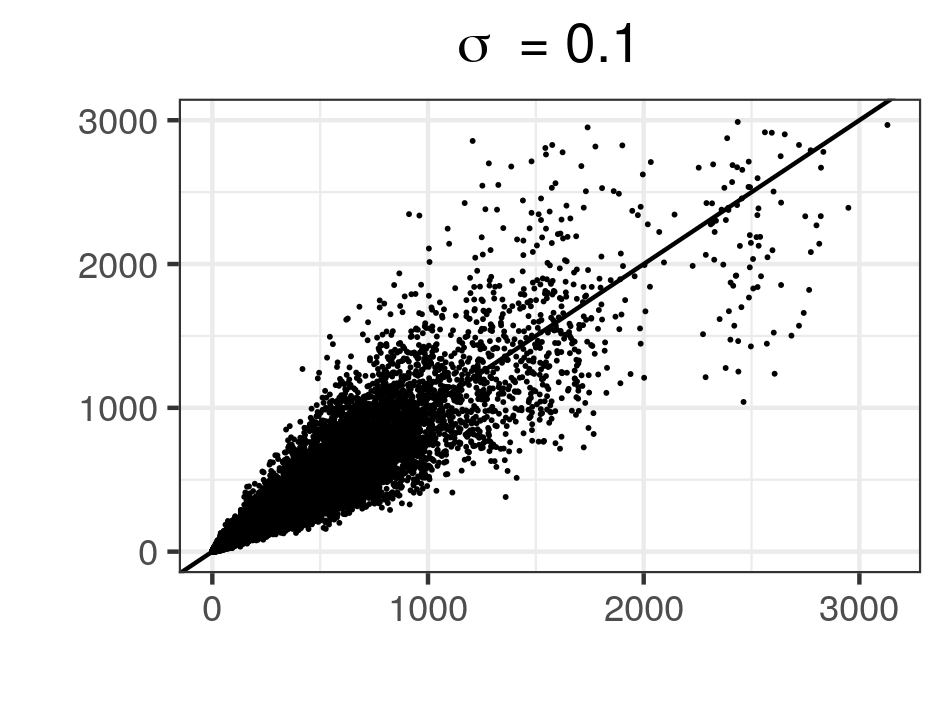}
\includegraphics{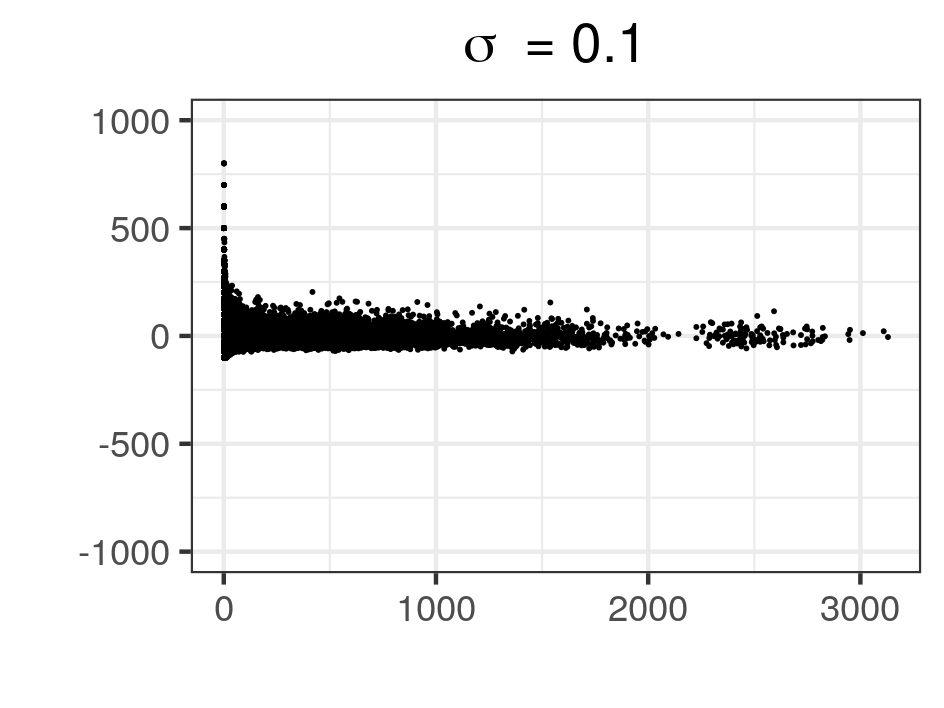}
\includegraphics{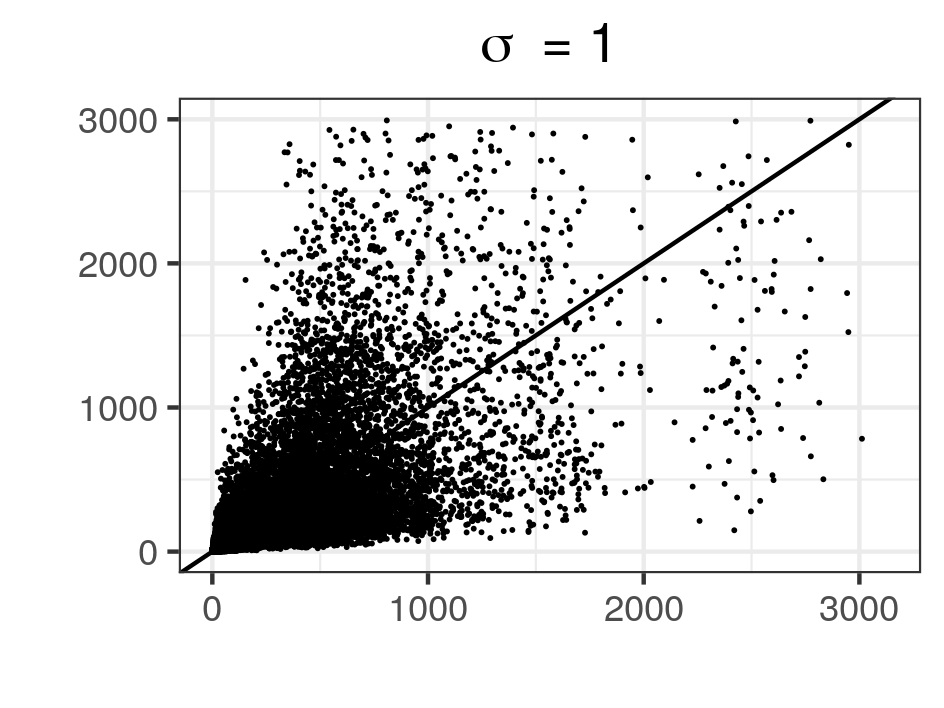}
\includegraphics{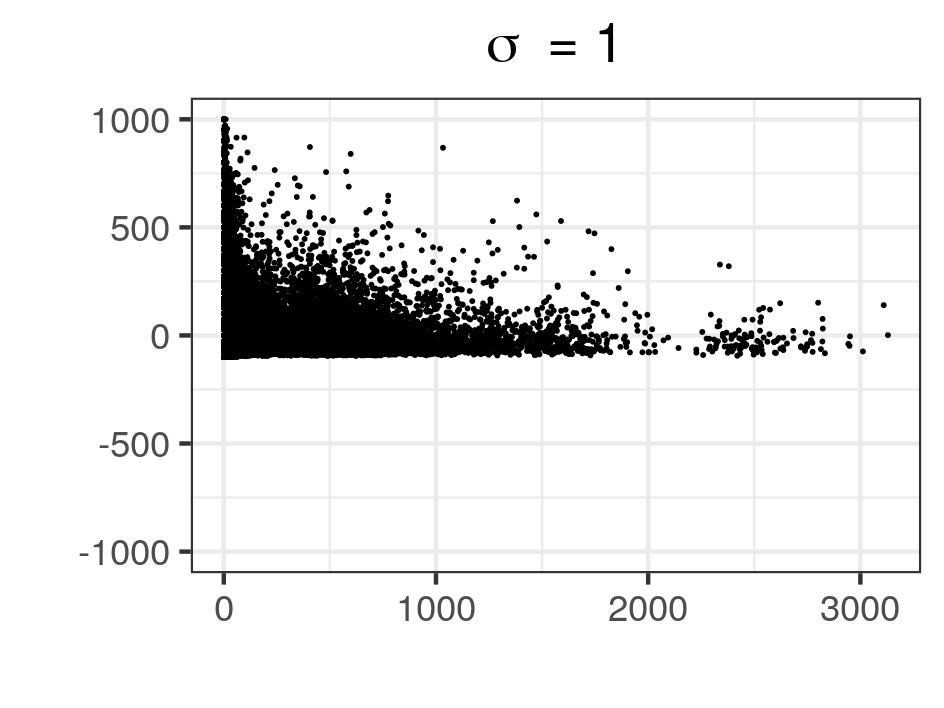}
\caption{\label{fig47} As Figure \ref{fig37} but for when the PIG was used rather than the NBI.}
\end{figure}

\begin{table}
\caption{\label{tab05}Empirical values obtained for the $\tau$ metrics for different $\sigma$ and $\alpha$ and for the NBI and PIG. }
\footnotesize
\centering
\fbox{
\begin{tabular}{*{11}{c}}
&& \multicolumn{4}{c}{NBI} &  & \multicolumn{4}{c}{PIG}  \\
\cline{3-6}
\cline{8-11}
& $k$ & 0 & 1 & 2 & 3 &   & 0& 1 & 2 & 3   \\
\cline{3-6}
\cline{8-11} \\
  & $\sigma$ & \multicolumn{9}{c}{$\tau_1(k)$} \\
  \multicolumn{1}{c|}{}&0 (Pois.) & 0.9190 & 0.0184 & 0.0135 & 0.0086  && 0.9190 & 0.0184 & 0.0135 & 0.0086  \\  
  \multicolumn{1}{c|}{} & 0.01 & 0.9191 & 0.0184 & 0.0134 & 0.0086  && 0.9192 & 0.0184 & 0.0134 & 0.0086  \\ 
 \multicolumn{1}{c|}{} &0.1 & 0.9204 & 0.0183 & 0.0130 & 0.0085  && 0.9203 & 0.0184 & 0.0130 & 0.0084  \\  
 \multicolumn{1}{c|}{$\alpha=0$}  &0.5 & 0.9256 & 0.0177 & 0.0117 & 0.0077  && 0.9243 & 0.0181 & 0.0121 & 0.0078  \\ 
 \multicolumn{1}{c|}{}  &1 &0.9317 & 0.0166 & 0.0105 & 0.0068  & & 0.9280 & 0.0179 & 0.0111 & 0.0072  \\ 
 \multicolumn{1}{c|}{}  &5 &0.9587 & 0.0098 & 0.0054 & 0.0035  && 0.9422 & 0.0167 & 0.0086 & 0.0053  \\
\multicolumn{1}{c|}{}  &10 & 0.9713 & 0.0064 & 0.0033 & 0.0022  && 0.9500 & 0.0156 & 0.0072 & 0.0042  \\   \\
\multicolumn{1}{c|}{}&0 (Pois.) &0.9013 & 0.0359 & 0.0136 & 0.0086  && 0.9013 & 0.0359 & 0.0136 & 0.0086  \\ 
\multicolumn{1}{c|}{} & 0.01 & 0.9012 & 0.0362 & 0.0136 & 0.0085  && 0.9013 & 0.0362 & 0.0136 & 0.0086  \\ 
\multicolumn{1}{c|}{} & 0.1 &0.9024 & 0.0360 & 0.0133 & 0.0084  && 0.9024 & 0.0361 & 0.0133 & 0.0085  \\ 
\multicolumn{1}{c|}{}  &0.5 &0.9078 & 0.0352 & 0.0120 & 0.0077  && 0.9065 & 0.0357 & 0.0122 & 0.0078  \\ 
 \multicolumn{1}{c|}{$\alpha=0.02$} &1 &0.9139 & 0.0339 & 0.0107 & 0.0069  && 0.9101 & 0.0355 & 0.0115 & 0.0072  \\ 
 \multicolumn{1}{c|}{} &5 & 0.9415 & 0.0259 & 0.0063 & 0.0036  && 0.9251 & 0.0330 & 0.0093 & 0.0053  \\ 
\multicolumn{1}{c|}{}  &10 &  0.9550 & 0.0212 & 0.0047 & 0.0024  && 0.9337 & 0.0307 & 0.0084 & 0.0045  \\ \\
  & & \multicolumn{9}{c}{$\tau_2(k)$} 
  \\ & & 0.9038 & 0.0346 & 0.0148 & 0.0075  &&  0.9038 & 0.0346 & 0.0148 & 0.0075  \\  \\
  & & \multicolumn{9}{c}{$\tau_3(k)$} \\
  \multicolumn{1}{c|}{}&0 (Pois.) & 1\;\;\;\;\;\;\;\; & 0.3674 & 0.2701 & 0.2231  & & 1\;\;\;\;\;\;\;\; & 0.3674 & 0.2701 & 0.2231  \\  
  \multicolumn{1}{c|}{} & 0.01 & 1\;\;\;\;\;\;\;\;& 0.3676 & 0.2706 & 0.2221  && 1\;\;\;\;\;\;\;\;& 0.3653 & 0.2703 & 0.2189  \\ 
 \multicolumn{1}{c|}{} &0.1 &1\;\;\;\;\;\;\;\; & 0.3489 & 0.2457 & 0.1976  & &1\;\;\;\;\;\;\;\; & 0.3538 & 0.2484 & 0.1974  \\ 
 \multicolumn{1}{c|}{$\alpha=0$}  &0.5 &  1\;\;\;\;\;\;\;\; & 0.2964 & 0.1874 & 0.1340  && 1\;\;\;\;\;\;\;\; & 0.3090 & 0.2022 & 0.1468  \\ 
 \multicolumn{1}{c|}{}  &1 &1\;\;\;\;\;\;\;\; & 0.2499 & 0.1489 & 0.1024  && 1\;\;\;\;\;\;\;\;& 0.2779 & 0.1677 & 0.1197  \\ 
 \multicolumn{1}{c|}{}  &5 &1\;\;\;\;\;\;\;\; & 0.1144 & 0.0618 & 0.0403  & &1\;\;\;\;\;\;\;\; & 0.1895 & 0.0981 & 0.0654  \\ 
\multicolumn{1}{c|}{}  &10 &  1\;\;\;\;\;\;\;\;& 0.0724 & 0.0378 & 0.0248  & &1\;\;\;\;\;\;\;\;& 0.1532 & 0.0740 & 0.0466  \\   \\
\multicolumn{1}{c|}{}&0 (Pois.) &0.9804 & 0.3648 & 0.2695 & 0.2247  && 0.9804 & 0.3648 & 0.2695 & 0.2247  \\ 
\multicolumn{1}{c|}{} & 0.01 &  0.9802 & 0.3645 & 0.2695 & 0.2186  && 0.9802 & 0.3656 & 0.2669 & 0.2217  \\ 
\multicolumn{1}{c|}{} & 0.1 &0.9802 & 0.3499 & 0.2450 & 0.1950  && 0.9801 & 0.3494 & 0.2466 & 0.1984  \\ 
\multicolumn{1}{c|}{}  &0.5 & 0.9803 & 0.2950 & 0.1876 & 0.1391  && 0.9803 & 0.3090 & 0.1981 & 0.1436  \\ 
 \multicolumn{1}{c|}{$\alpha=0.02$} &1 &0.9804 & 0.2515 & 0.1498 & 0.1052  && 0.9803 & 0.2782 & 0.1689 & 0.1218  \\ 
 \multicolumn{1}{c|}{} &5 &0.9812 & 0.1172 & 0.0620 & 0.0429  && 0.9810 & 0.1888 & 0.0959 & 0.0629  \\ 
\multicolumn{1}{c|}{}  &10 & 0.9819 & 0.0711 & 0.0374 & 0.0255  && 0.9817 & 0.1524 & 0.0750 & 0.0486  \\  \\
  & & \multicolumn{9}{c}{$\tau_4(k)$} \\
  \multicolumn{1}{c|}{}&0 (Pois.) & 0.9835 & 0.6893 & 0.2974 & 0.1943  && 0.9835 & 0.6893 & 0.2974 & 0.1943  \\   
  \multicolumn{1}{c|}{} & 0.01 & 0.9834 & 0.6890 & 0.2995 & 0.1938  && 0.9833 & 0.6867 & 0.2989 & 0.1905  \\ 
 \multicolumn{1}{c|}{} &0.1 &  0.9820 & 0.6603 & 0.2811 & 0.1742  && 0.9821 & 0.6648 & 0.2827 & 0.1760  \\  
 \multicolumn{1}{c|}{$\alpha=0$}  &0.5 & 0.9764 & 0.5788 & 0.2372 & 0.1304  && 0.9778 & 0.5890 & 0.2484 & 0.1416  \\  
 \multicolumn{1}{c|}{}  &1 &0.9701 & 0.5203 & 0.2108 & 0.1125  && 0.9739 & 0.5369 & 0.2232 & 0.1243  \\ 
 \multicolumn{1}{c|}{}  &5 &0.9427 & 0.4043 & 0.1710 & 0.0858  && 0.9593 & 0.3919 & 0.1694 & 0.0929  \\ 
\multicolumn{1}{c|}{}  &10 & 0.9305 & 0.3910 & 0.1677 & 0.0851  && 0.9513 & 0.3387 & 0.1521 & 0.0822  \\   \\
\multicolumn{1}{c|}{}&0 (Pois.) &0.9831 & 0.3516 & 0.2935 & 0.1957  && 0.9831 & 0.3516 & 0.2935 & 0.1957  \\ 
\multicolumn{1}{c|}{} & 0.01 & 0.9829 & 0.3484 & 0.2935 & 0.1919  && 0.9830 & 0.3495 & 0.2911 & 0.1934  \\ 
\multicolumn{1}{c|}{} & 0.1 &0.9817 & 0.3357 & 0.2735 & 0.1733  && 0.9817 & 0.3350 & 0.2745 & 0.1751  \\ 
\multicolumn{1}{c|}{}  &0.5 &0.9759 & 0.2898 & 0.2316 & 0.1348  && 0.9774 & 0.2991 & 0.2403 & 0.1379  \\ 
 \multicolumn{1}{c|}{$\alpha=0.02$} &1 &0.9696 & 0.2567 & 0.2066 & 0.1141 && 0.9735 & 0.2712 & 0.2168 & 0.1259  \\ 
 \multicolumn{1}{c|}{} &5 & 0.9419 & 0.1562 & 0.1469 & 0.0890  && 0.9584 & 0.1979 & 0.1520 & 0.0887  \\ 
\multicolumn{1}{c|}{}  &10 & 0.9293 & 0.1162 & 0.1172 & 0.0799  && 0.9503 & 0.1715 & 0.1320 & 0.0808 
\end{tabular}}
\end{table}


\subsection{Testing {specific utility}  through log-linear model analysis}
In the synthetic data literature, specific utility \citep{Snoke2018} is often assessed by comparing inferences, such as regression coefficients, obtained from the original and synthetic data. \par
The synthesizer does not know{,} of course{,} what analyses users of the synthetic data would perform. Among the variables included in the data, which are best described as demographic, there is no obvious response variable, so analysts may be interested in associations between variables. {Therefore, a log-linear analysis was chosen as a suitable way to test the specific utility of synthetic data generated with the synthesis method described in Section 3.} It is difficult to obtain parameter estimates and parameters' standard error estimates for the full five-variable data, since large amounts of memory and storage are required {- the same problem faced when fitting synthesis models}. To relieve some of this pressure, the all two-way interaction model was fitted to three of the data's five variables, ethnicity, age and language, resulting in 608 parameters. 
\par
The confidence interval overlap metric \citep{Karr2006} was used to measure similarities between estimates. {In order to define confidence interval overlap, let $(l_o, u_o)$ and $(l_s, u_s)$ denote confidence intervals for a univariate population parameter $Q$, obtained from the original and synthetic data, respectively; and let $(l_i, u_i)$ denote the intersection  of the two intervals, that is, $l_i=\text{max}(l_o, l_s)$ and $u_i=\text{min}(u_o, u_s)$. Then the confidence interval overlap $I_Q$ is given as:}
\begin{align}
I_Q = \frac{1}{2} \bigg( \frac{u_i-l_i}{u_o-l_o} + \frac{u_i-l_i}{u_s-l_s} \bigg). \label{overlap}
\end{align} 
{Thus $I_Q$ is the mean of two ratios: the length of the confidence interval intersection divided by (i) the length of the confidence interval from the original data, and (ii) the length of the confidence interval from the synthetic data.} \par
{Combining rules are required to obtain valid parameter estimates and standard errors from synthetic data, even when just $m=1$ synthetic data set is generated. This is because there are {always} two sources of uncertainty in synthetic data that need to be accounted for: the sampling uncertainty inherent in the original data, and the uncertainty owing to synthesis. To simplify the analysis - after all, the purpose here is just to evaluate the utility of the synthetic data - the original data were assumed to constitute a simple random sample drawn from a super-population. This allowed the estimator given in \cite{Raab2016} to be used, which provides valid variance estimates for large samples when analysing synthetic data generated through the mechanism described in Section \ref{section3}. When estimating a population parameter $Q$ from $m\geq1$ synthetic data sets, $\hat{Q}$ is found by averaging over the $m$ data sets, and its variance is given as:}}
\begin{align}
\widehat{\text{Var}}(\hat{Q})&= \bar{v}_m ( {n_\text{syn}}/{n} + {1}/{m}), \label{Raabestimator}
\end{align}
{where $\bar{v}_m$ is the mean variance estimate across the $m$ synthetic data sets, and $n_\text{syn}$ and $n$ are the ``sample'' sizes of the synthetic and original data, respectively. {Unlike other estimators, such as the one given in \cite{Reiter2003}, this estimator allows valid variance estimates to be obtained from just $m=1$ synthetic data set, as done here. When $m=1$ ($\bar{v}_m=v$) and $n=n_\text{syn}$, the estimator in (\ref{Raabestimator}) simplifies to $2v$, that is, the variance estimate from the synthetic data, doubled.  } 
\par {Finally, in log-linear models,} estimability issues {can arise} through the presence of zero counts in the data. This {can lead} to issues surrounding non-existence and non-identifiability of estimates \citep{Fienberg2012}{. But} no serious model fitting issues arose {in this particular example. T}here {we}re some parameters included in the model with a true value of $-\infty$. For such parameters, \proglang{R} returned a large negative value, typically in the vicinity of -20.
 \subsubsection{Results}
\begin{table}
\caption{\label{table6} How $\sigma$ and $\alpha$ affect the trimmed mean (top and bottom $10\%$ excluded) percentage difference between log-linear parameter estimates obtained from the observed and synthetic data. The trimmed mean was used to subdue the effect of huge percentage differences arising through the presence of zero counts. For clarity, for an arbitrary original log-linear parameter estimate $q$ and its corresponding synthetic estimate $q^\text{syn}$, the percentage difference was calculated by $100 \times (q^\text{syn}-q)/q$.}
\centering
\fbox{ \begin{tabular}{*{8}{lrrrrrrr}}
& \em $\sigma=0$ & \em $\sigma=0.1$ & \em $\sigma=0.5$ & \em $\sigma=1$ & \em $\sigma=2$& \em $\sigma=5$& $\sigma=10$\\
& (Pois.) &  & \em  & \em  & \em& \em& \\
\hline \\
\multicolumn{8}{l}{\textbf{The NBI model}} \\
$\alpha=0$ & -1.7 & 3.9 & -12.0 & -0.1 & -18.7 & 10.6 & -108.4 \\ 
  $\alpha=0.005$  & -23.5 & -30.9 & -31.7 & -34.8 & -34.0 & 14.8 & -32.9 \\ 
  $\alpha=0.01$  & -32.5 & -33.2 & -33.7 & -38.9 & -47.8 & 41.0 & -64.7 \\ 
  $\alpha=0.015$  & -38.8 & -39.8 & -47.7 & -52.3 & -47.6 & -20.3 & -3.5 \\ 
  $\alpha=0.02$  & -37.1 & -34.0 & -44.4 & -40.9 & -27.7 & -42.0 & -33.5 \\ \\
  \multicolumn{8}{l}{\textbf{The PIG model}} \\
  $\alpha=0$  & -1.7 & -2.2 & 16.2 & 11.2 & -33.0 & -6.6 & 187.4 \\ 
  $\alpha=0.005$  & -23.5 & -21.7 & -28.4 & -21.7 & -33.0 & -73.6 & -990.4 \\ 
  $\alpha=0.01$  & -32.5 & -29.6 & -31.7 & -48.6 & -42.3 & 25.9 & -399.3 \\ 
  $\alpha=0.015$  & -38.8 & -26.4 & -36.6 & -37.6 & -20.6 & -64.4 & -504.0 \\ 
  $\alpha=0.02$  & -37.1 & -47.8 & -40.2 & -20.6 & -40.5 & -76.2 & 425.2 \\ 
\end{tabular} }
\end{table}
Figures \ref{fig1} and \ref{fig6} present boxplots of confidence interval overlap values for the the log-linear model {parameters} across {the} different synthesis models. 
They demonstrate how increasing $\sigma$ and $\alpha$ causes utility to fall away. For example, irrespective of $\alpha$, whenever $\sigma=10$, the median confidence interval overlap is zero. {In general, a} high proportion of the overlap values are equal to 1/2. {This can be seen, for example, in the centre and right plots of Figure \ref{fig6}, where several of the upper quartiles are equal to 1/2.} This{, incidentally,} is owing to the nature - and perhaps a criticism - of the confidence interval overlap metric{ (given in \ref{overlap}): w}henever {one of the} confidence interval{s' lengths} tends {to} infinit{y - but the other confidence interval is finite - the overlap value tends to 1/2.} 

Table \ref{table6} presents (trimmed) mean percentage differences between synthetic and observed parameter estimates for {various $\sigma$ and $\alpha$}. {Even s}etting $\alpha$ small can have an adverse effect on utility. For example, when $\sigma=0$ (the Poisson model), increasing $\alpha$ from 0 to 0.005 causes the (trimmed) mean percentage difference {in estimates} to {fall} from -1.7$\%$ to $-23.5\%$, thus demonstrating the bias {caused by} $\alpha>0$. {T}he general trend {is} that increasing $\alpha$ and $\sigma$ result{s} in larger percentage differences.
\subsection{Balancing risk and utility}
A key question a synthesizer would have is: which synthesis method offers the best balance between utility and risk? To address this, the risk-utility trade{-}off from each generated synthetic data set can be plotted. An example is displayed in Figure \ref{fig15}. Privacy has been measured on the y-axis via $1-\tau_4(1)$, that is, (1 - risk), and utility on the x-axis by mean confidence interval overlap. The original data sit at the point (1,0), that is, maximum utility and minimum privacy. All points must lie with{in} the unit square $[0,1]\times[0,1]$ and the further from the origin, the better the synthetic data. {For instance, when a point is near the origin, it suggests that for the same level of privacy, a greater level of utility is achievable (or vice versa).}

This visualisation offers a convenient way to compare the performance of different synthesis {models}. For example, it may be possible {for} one synthetic data set {to strictly dominate} another: the PIG model with $\sigma=10$ provides greater utility and lower risk than the NBI model with $\sigma=100$. {T}he choice depends on the priorities of the data holder and users. For example, it may be that synthetic data can only be released if $\tau_4(1)$ is at least 0.5, in which case the synthetic data with the highest utility that satisfies this requirement could be released, here this would be the PIG model with $\sigma = 10$. Alternatively, it may be that only data with a utility value of at least 0.5 would be deemed useful enough for release, in which case the synthetic data generated under a NBI model with $\sigma = 0.1$ would be chosen.

In practice, a range of different metrics for utility and privacy can be created and feed into determining which synthesis method {is chosen}. This decision is also likely to be application specific.

\begin{figure}
\centering
\makebox{\includegraphics[height=5cm]{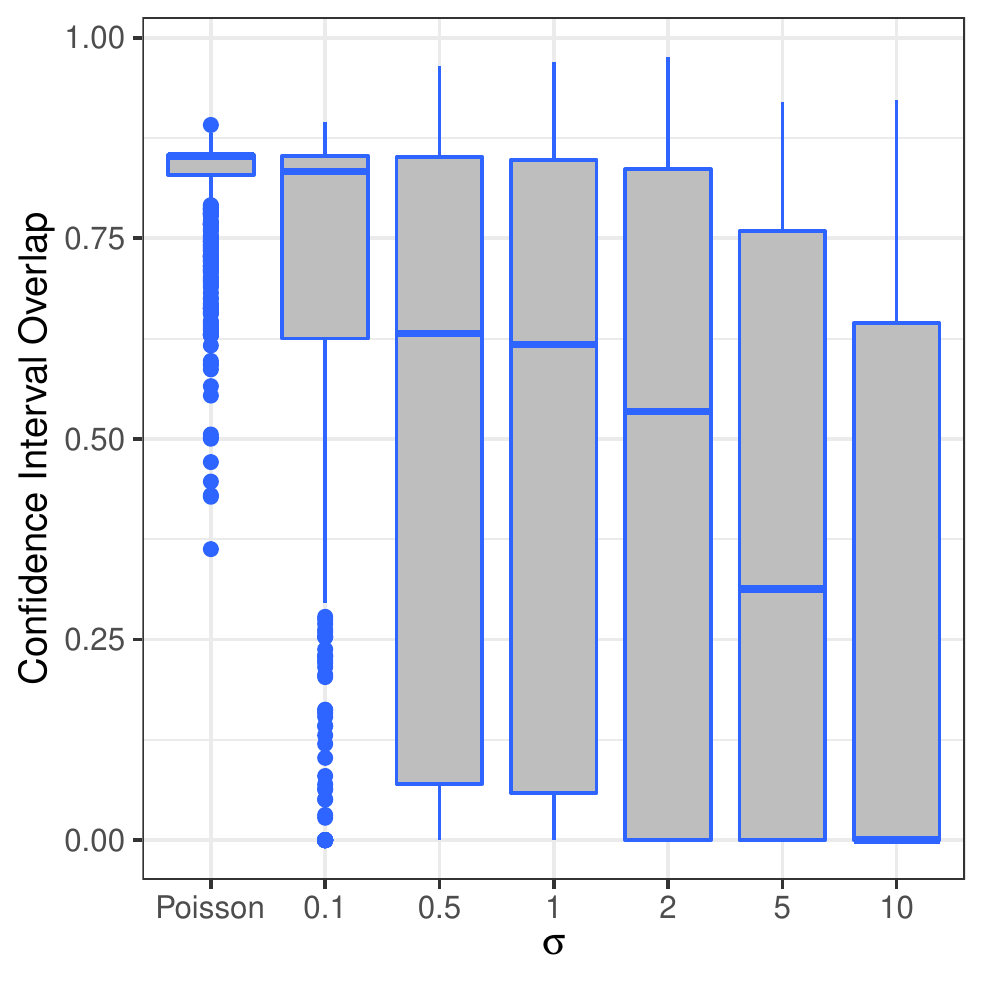}}
\makebox{\includegraphics[height=5cm]{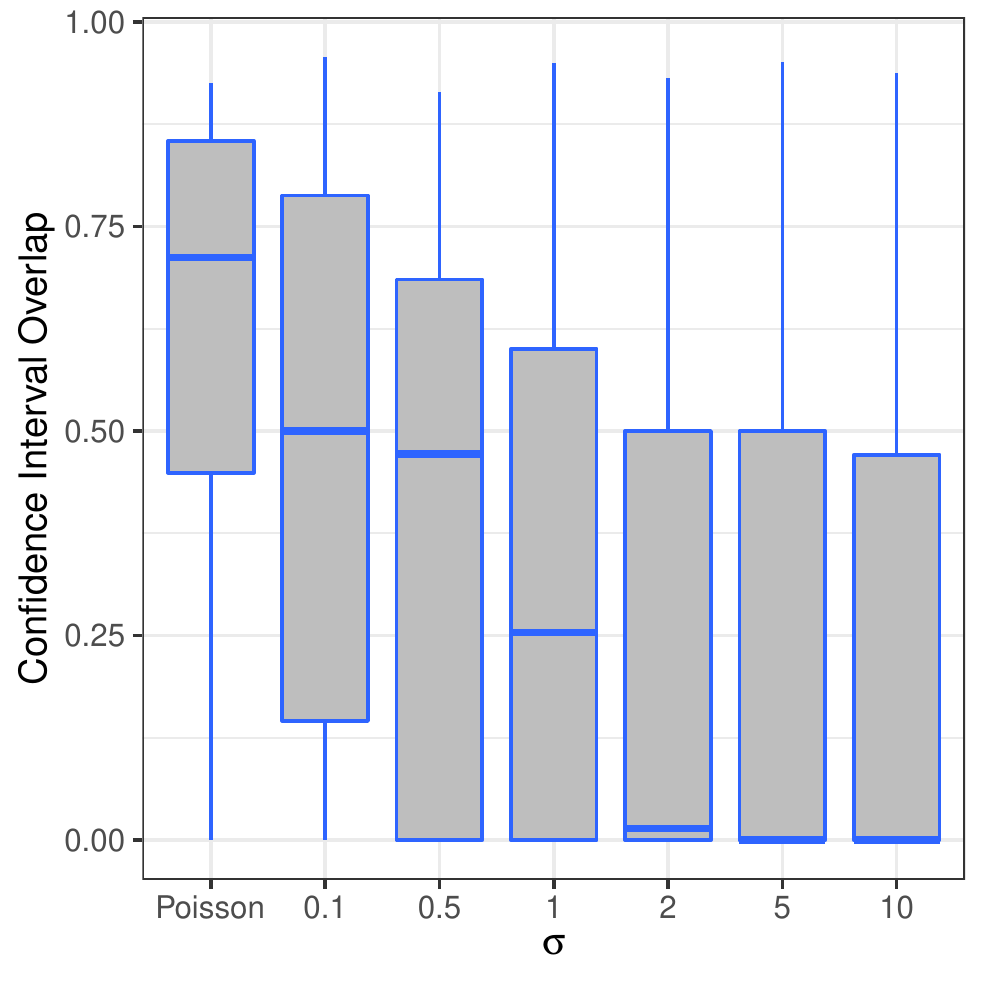}}
\makebox{\includegraphics[height=5cm]{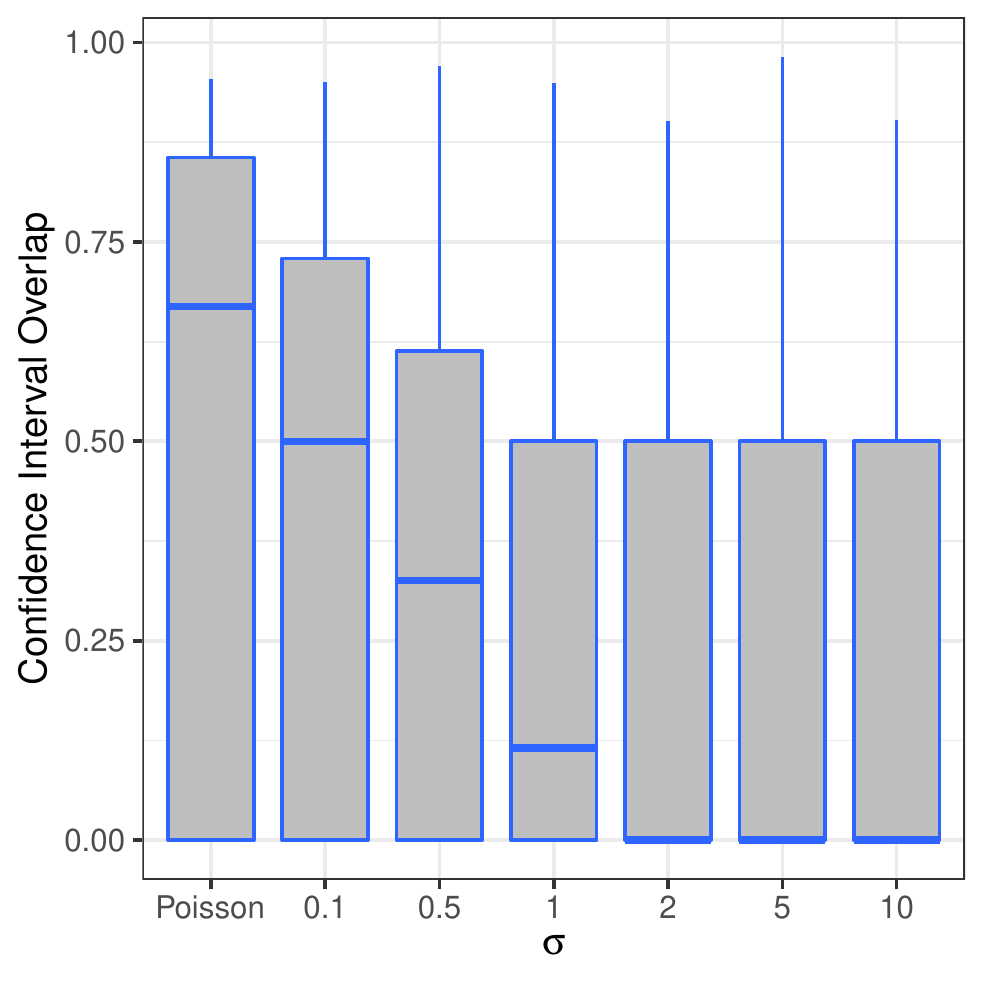}}
\caption{\label{fig1} These boxplots show how $\sigma$ and $\alpha$ affect log-linear parameters' confidence interval overlap when the NBI distribution is used for synthesis. The left frame is the case where $\alpha=0$; the middle frame where $\alpha=0.01$; and the right frame where $\alpha=0.02$.}
\end{figure} 
\begin{figure}
\centering
\makebox{\includegraphics[height=5cm]{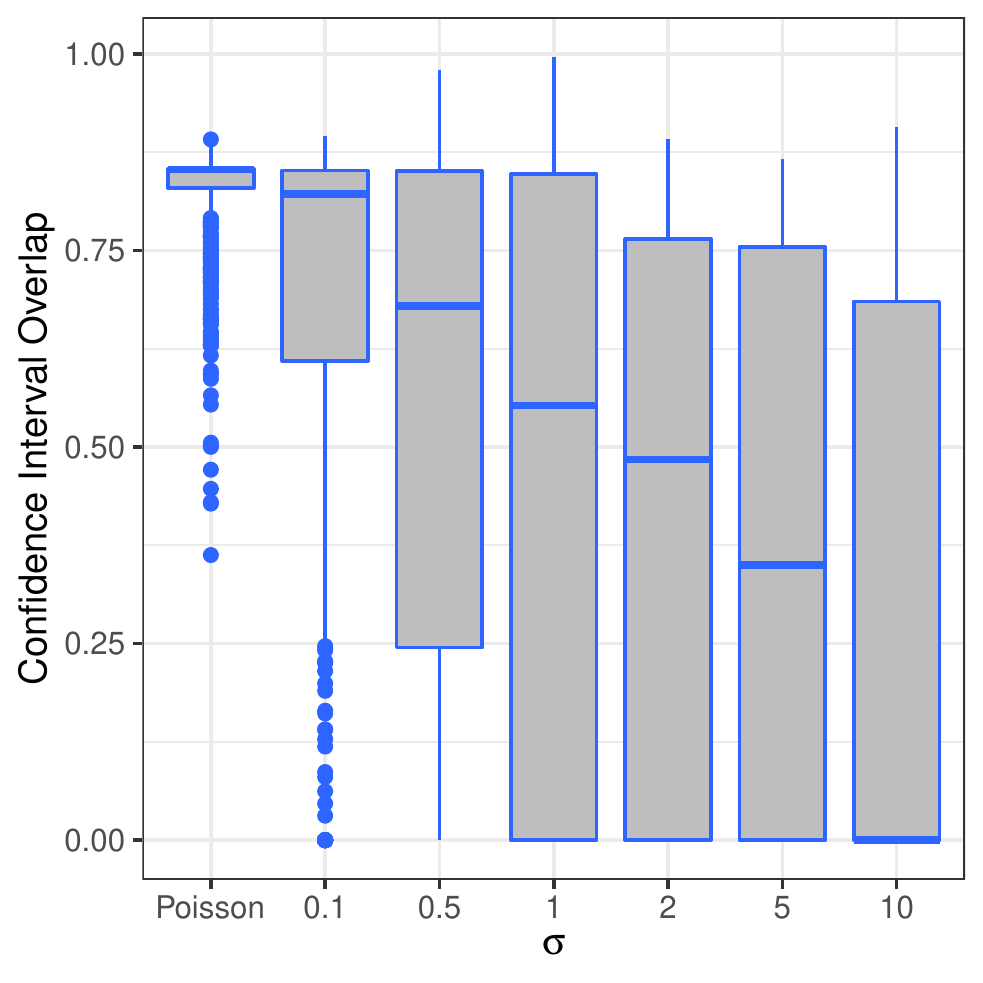}}
\makebox{\includegraphics[height=5cm]{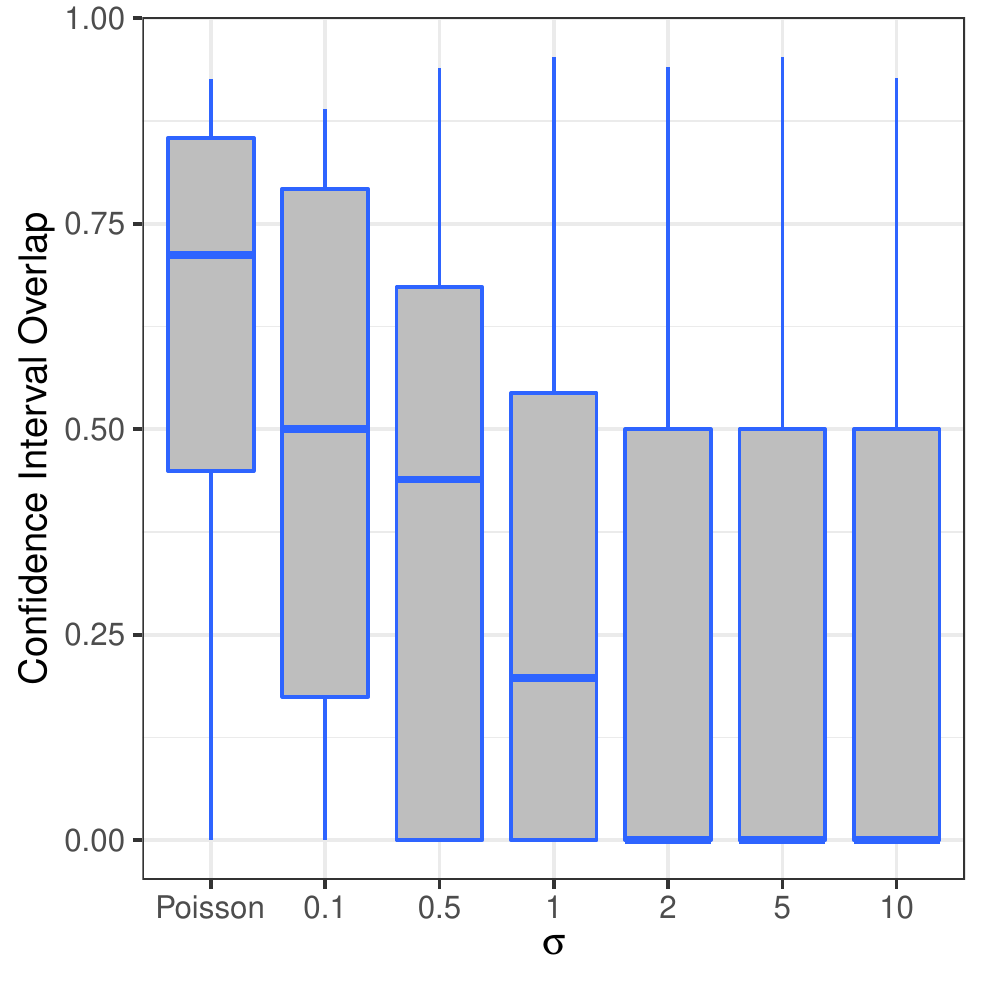}}
\makebox{\includegraphics[height=5cm]{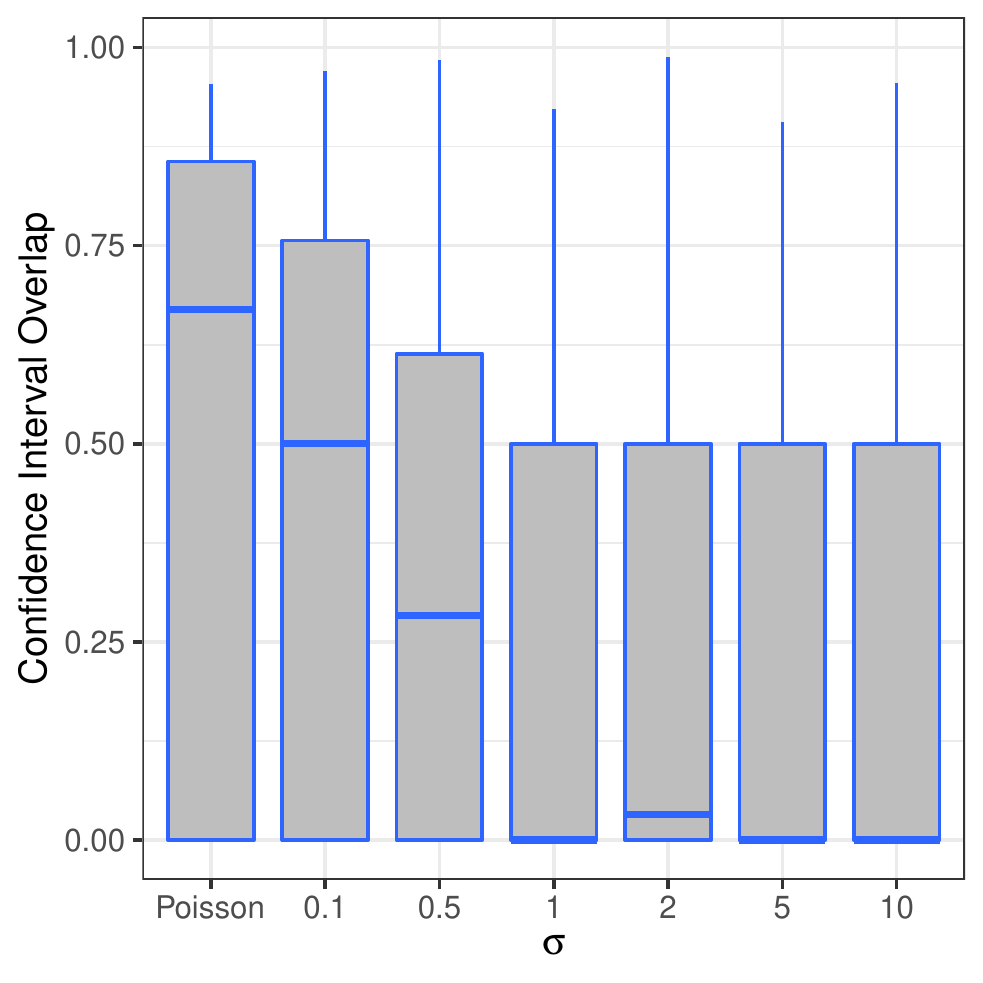}}
\caption{\label{fig6} These boxplots show how $\sigma$ and $\alpha$ affects confidence interval overlap when the PIG distribution is used for synthesis. The left frame is the case where $\alpha=0$; the middle frame where $\alpha=0.01$; and the right frame where $\alpha=0.02$.}
\end{figure} 

\begin{figure}
\centering
\centering
\makebox{\includegraphics[width=\textwidth]{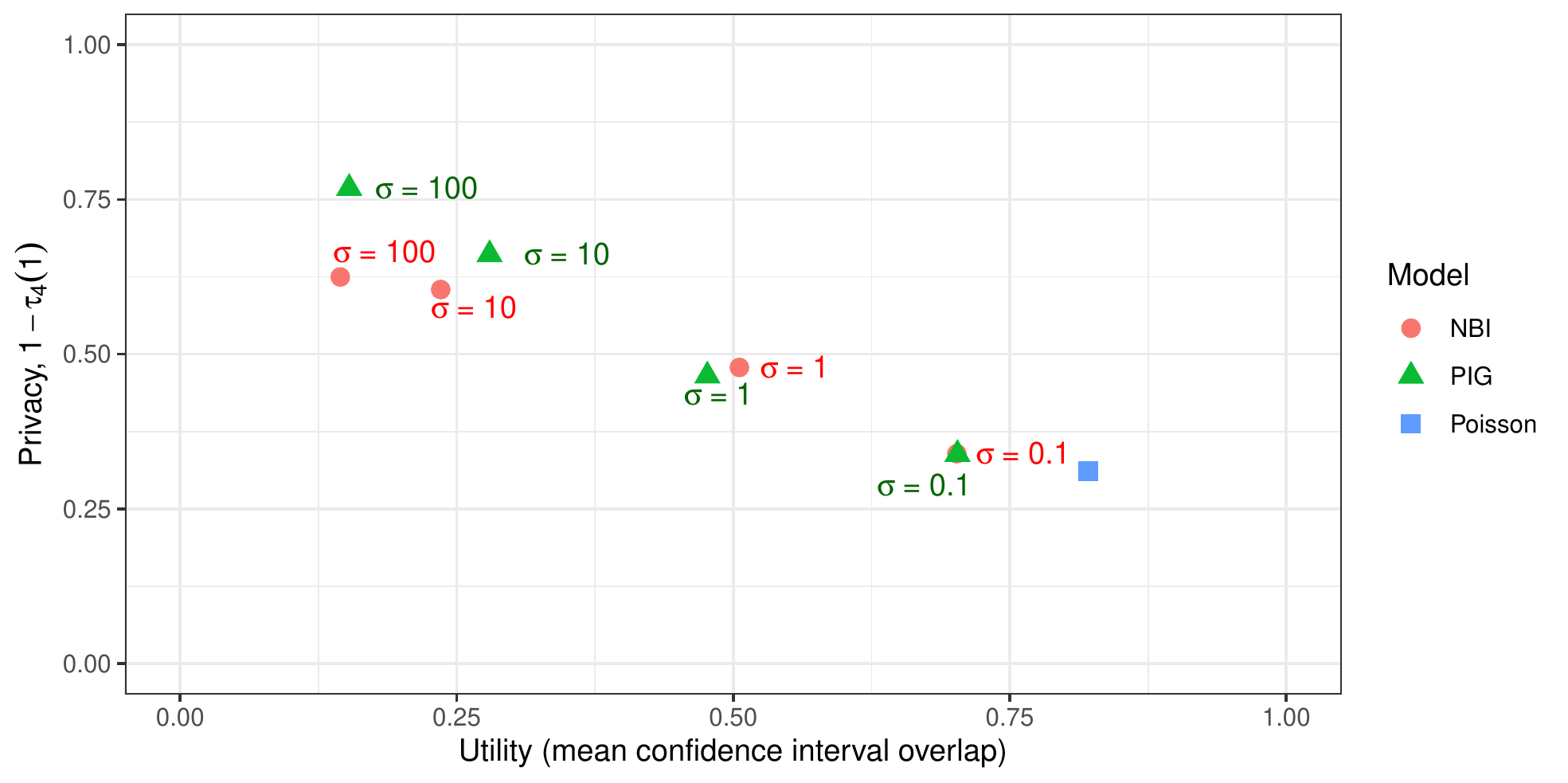}}
\caption{\label{fig15} This plot, which is resemblant of a product possibility frontier in economics, provides a visual representation of the risk-utility trade-off for different $\sigma$ ($\alpha=0$).}
\end{figure} 

\subsection{The Poisson model versus the NBI model versus the PIG model} 
{
The intention is that the synthesizer would usually use the NBI or PIG distributions, rather than the Poisson, which is far too limited to be used in practice. In general, the NBI and PIG models give similar results, yet this is to be expected as both share the same variance function. Nevertheless, there are some marked differences between the two, especially when $\sigma$ is large; for example, when $\sigma=10$ and $\alpha=0$, there are substantially fewer zeros in the synthetic data when the PIG is used than when the NBI is used ($\tau_1(0)$ values of 0.950 and 0.971, respectively). There is, of course, scope to use other count distributions here; those with a different variance function would have an entirely different profile altogether. Moreover, both the NBI and PIG are also both limited in that they can only model overdispersion and not underdispersion, hence the variance is always greater than the Poisson - and this may be unnecessary. The double Poisson distribution (see \citealt{Rigby2019}), for example, which can be used to model underdispersed count data, would allow the variance to be set lower than in the Poisson.}
\section{Discussion and future work}


{In this paper, the case of generating $m=1$ synthetic data set was considered. But $m>1$ data sets can be generated using the same framework and, in the same way, certain properties can be found analytically. There might be advantages of doing this, especially in relation to the risk-utility trade-off. It effectively introduces another tuning parameter, thus providing further flexibility. Investigations unreported in this paper have shown promising signs for $m>1$; for example, particularly for large $\sigma$, the gains in utility appear to outweigh the relatively small increase in risk. Moreover, an optimal value for $n_\text{syn}$, the sample size of the synthetic data, was not sought. There is scope to set $n_\text{syn}$ lower or higher than $n$ (the sample size of the original data), which again can be evaluated in relation to the risk-utility trade-off. \par
While the two-parameter count distributions allow the synthesizer to set the synthetic counts' variance, they cannot control where the variability falls. It is not desirable for the variability to manifest itself in, say, a heav{y} right tail {in the synthesis distribution's probability mass function}, because, while some movement is required in synthetic counts to reduce risk, large movements are unnecessary and may have} an adverse effect on the data's utility. The use of three-parameter count distributions, such as the Delaporte and Sichel distributions (see \citealt{Rigby2019}), would provide the synthesizer with control over the skewness in addition to the variance. 

{This method assumes that the cell counts in the multi-way table are independent. This assumption} can be {exploited} further by specifying different models - {for example,} different $\sigma$ and $\alpha$ values - {when synthesizing} different {parts of} the original data{'s multi-way table}. Smaller cell sizes could be synthesized using a relatively larger $\sigma$ than larger cells, which would inject more variability where it is needed. \par {The method as presented here also assumes that the size of $\alpha$ is constant across all random zeros. However, it can be argued that some random zeros in the original data are more (or less) likely to be non-zeros than others; for example, some zero cells pertain to higher order marginal counts that are also zero. This can be accounted for, to a certain extent, by smoothing the original counts through fitting, for example, an all two-way interaction log-linear model. But the benefits of using saturated models would be lost. The pseudo-Bayes estimator, as presented in Chapter 12 of \cite{Bishop1975}, provides an alternative to adding constant $\alpha$. A set of prior cell probabilities (denoted by $\boldsymbol\lambda$) are selected using, for example, external information. Based on these prior probabilities, {the observed counts are re-weighted to provide a set of adjusted counts, and a saturated model would then be applied as before, synthesizing from these adjusted values}. {Hence, while the observed counts are smoothed - and potentially reducing the number of zero cells, thereby helping to minimise the impact of the problem discussed in Section \ref{addsmooth} - they are not smoothed through modelling decisions (setting interactions to zero), but through the choice of $\boldsymbol\lambda$.} {This means, however, that the fundamental challenge} is just transferred from choosing $\alpha$ to choosing $\boldsymbol\lambda$. The strategies provided in \cite{Bishop1975} may offer some insights into this, although the objectives of the synthesis would also be relevant here. This is something that would involve further careful consideration and is a substantial research question on its own.}

{In the empirical study (Section 4), only variables at the pupil-level were considered. However, administrative data might have a hierarchical structure that would need to be taken into account. In this example, this could involve incorporating school-level variables into the synthesis. This clearly presents challenges from the modelling and utility perspective, such as ensuring relationships between pupil-level variables within schools are preserved in the synthetic data. However, this also presents interesting questions around disclosure risk because the school-level variables may increase the risk at the pupil-level. The risk and utility challenges associated with multi-level data in this area merit further consideration.}

\par There is no panacea for synthetic data generation. A compromise always needs to be struck between risk, utility and, in the case of large data sets, {computational} time. {Different methods are, of course, suited to different data types and sizes. The conditional approaches outlined in Section \ref{conditional}, which typically either use GLMs or CART, are effective in synthesizing microdata sets, particularly those comprising a mix of continuous and categorical variables. Yet, when $n$ is large, demands on memory makes it challenging, computationally, to implement such methods. It is more efficient to undertake synthesis of categorical data at the aggregated level. Among these approaches, the advantage of using saturated models is two-fold. They eliminate the need to make modelling decisions, which ensures the preservation of relationships, and support an \textit{a priori} approach to synthesis, whereby expected properties of the synthetic data can be established beforehand. This does not just apply to the ``$\tau$ metrics'' used in this paper, but to many other risk and utility metrics; for example, expressions can be similarly derived - or at least approximated - for general utility measures such as Hellinger distance and Kulback-Leibler divergence. This facilitates a more formal approach to risk and utility that may, in turn, invite greater transparency. } \par This paper hopefully gives confidence to organisations holding large administrative databases that generating synthetic data is not necessarily a computationally intensive and time{-}consuming endeavour. Further, the organisations can easily tune the synthesis models in a very transparent way to achieve pre-specified levels of risk and utility. 
\section*{Acknowledgements}
This work was carried out as part of a CASE PhD Studentship between Lancaster University and the Office for National Statistics, funded by the Economic and Social Research Council (ESRC). 
\bibliography{FINAL_arxiv.bib}
\end{document}


\maketitle

In this supplement we derive expressions relevant to the negative binomial (NBI) and Poisson-inverse Gaussian synthesis models presented in the main body. Section 1 presents expressions for the $\tau$ metrics, while Sections 2 and 3 derive values for $\alpha$ and $\sigma$ that satisfy certain properties which may be of interest to the synthesizer.

\section{The $\tau$ metrics for the NBI and PIG models}
\subsection{The metrics $\tau_1$ and $\tau_2$}
The metric $\tau_1(k)$ is the proportion of cells of size $k$ in the synthetic data and $\tau_2(k)$ is the proportion of cells of size $k$ in the original data,
\begin{align}
\tau_1(k)&=p(f^{\text{syn}}=k), \nonumber 
\intertext{which, by the law of total probability,} \nonumber
&=   p(f^{\text{syn}}=k \mid f=0, \;\alpha) \cdot p(f=0) +  \sum\limits_{j=1}^\infty p(f^{\text{syn}}=k \mid f=j) \cdot p(f=j). \nonumber 
\intertext{\subsubsection{The negative binomial (NBI)}
When the NBI model is used for synthesis,}
\tau_1(k)&= \frac{\Gamma(k+{1}/{\sigma})}{\Gamma(k+1)\Gamma({1}/{\sigma})} \cdot \frac{(\sigma \alpha)^{k}}{(1+\sigma \alpha)^{k+1/\sigma}} \cdot \tau_2(0) + \sum\limits_{j=1}^\infty \frac{\Gamma(k+{1}/{\sigma})}{\Gamma(k+1) \Gamma({1}/{\sigma})} \cdot \frac{(\sigma j)^{k}}{(1+\sigma j)^{k+1/\sigma}}  \cdot \tau_2(j) . \label{tau1NBI} 
\intertext{\subsubsection{The Poisson-inverse Gaussian (PIG)}
When the PIG model is used for synthesis,}
\tau_1(k) &=  \bigg(\frac{2c_\alpha}{\pi}\bigg)^{1/2} \cdot \frac{ \alpha^{k} \text{exp}({1/\sigma}) K_{k-1/2}(c_\alpha)}{(c_\alpha \sigma)^{k} k!} \cdot \tau_2(0) + \sum\limits_{j=1}^\infty \bigg(\frac{2c_j}{\pi}\bigg)^{1/2} \cdot \frac{ j^{k} \text{exp}({1/\sigma}) K_{k-1/2}(c_j)}{(c_j \sigma)^{k} k!} \cdot \tau_2(j), \nonumber \\
&\text{where} \quad c_j^2=\frac{1}{\sigma^2}+\frac{2 j}{\sigma} \quad
\text{and} \quad K_{\lambda}(t)=\frac{1}{2}\int\limits_0^{\infty} x^{\lambda-1} \exp\bigg\{-\frac{1}{2}t(x+x^{-1})\bigg\} \;\text{d}x \label{tau0PIG}
\intertext{is the modified Bessel function of the third kind.
\subsection{The metric $\tau_3$}
The metric $\tau_3(k)$ is the proportion of cells of size $k$ which are synthesized to $k$,}
\tau_3(k)&=p(f^{\text{syn}}=k | f	=k). \nonumber 
  \intertext{\subsubsection{The negative binomial (NBI)}}
\tau_3(k)&=  
\begin{cases}
\displaystyle
   \frac{1}{(1+\sigma \alpha)^{1/\sigma}} & \text{if $k=0$ } \\ \\
   \displaystyle
     \frac{\Gamma(k+{1}/{\sigma})}{\Gamma(k+1) \cdot \Gamma({1}/{\sigma})} \cdot \frac{(\sigma k)^k}{(1+\sigma k)^{k+1/\sigma}} & \text{if $k\geq1$} 
  \end{cases} \nonumber
  \intertext{\subsubsection{The Poisson-inverse Gaussian (PIG)}}
  \tau_3(k)&=  
\begin{cases}
\displaystyle
   \text{exp}({1/\sigma}-c_\alpha) & \text{if $k=0$ } \\ \\
   \displaystyle
     \bigg(\frac{2c_k}{\pi}\bigg)^{1/2} \cdot \frac{ k^{k} \text{exp}({1/\sigma}) K_{k-1/2}(c_k)}{(c_k \sigma)^{k} k!}  & \text{if $k\geq1$}
  \end{cases} \nonumber
\intertext{and $c_j^2$ is given in (\ref{tau0PIG}).
\subsection{The metric $\tau_4$}
The metric $\tau_4(k)$ is the proportion of cells of size $k$ in the synthetic data which originated from a cell of size $k$,} 
\tau_4(k)&= p(f=k | f^{\text{syn}}	=k) = \frac{p(f^{\text{syn}}=k | f	=k)p(f	=k) }{p(f^\text{syn}=k)}= \frac{\tau_3(k)\tau_2(k)}{\tau_1(k)}. \nonumber
  \intertext{\subsubsection{The negative binomial (NBI)}}
  \tau_4(k)&=  
\begin{cases}
\displaystyle
   {\frac{1}{(1+\sigma \alpha)^{1/\sigma}} \cdot \tau_2(0)}\bigg/ \bigg({\frac{1}{(1+\sigma \alpha)^{1/\sigma}} \cdot \tau_2(0) + \sum\limits_{j=1}^\infty \frac{1}{(1+\sigma j)^{1/\sigma}} \cdot \tau_2(j)\bigg)} & \text{if $k=0$ } \\ \\
   \displaystyle
     \frac{k^k}{(1+k\sigma)^{k+1/\sigma}} \cdot \tau_2(k) \bigg/\bigg( \frac{\alpha^k}{(1+\alpha\sigma)^{k+1/\sigma}} \cdot \tau_2(0) + \sum\limits_{j=1}^\infty \frac{j^k}{(1+j\sigma)^{k+1/\sigma}} \cdot \tau_2(j) \bigg) & \text{if $k\geq1$}  \label{NBItau4}
  \end{cases}
\intertext{after cancelling constant terms.} \nonumber
 \intertext{\subsubsection{The Poisson-inverse Gaussian (PIG)}}
   \tau_4(k)&=  
\begin{cases}
\displaystyle
   \text{exp}(-c_\alpha) \cdot \tau_2(0) \bigg/\bigg(\text{exp}(-c_\alpha) \cdot \tau_2(0) + \sum\limits_{j=1}^\infty{ \text{exp}(-c_j)} \cdot \tau_2(j) \bigg)  & \text{if $k=0$ } \\ \\
   \displaystyle
     {c_k}^{\frac{1}{2}-k}  { k^{k} K_{k-\frac{1}{2}}(c_k)} \cdot \tau_2(0)  \bigg/ \bigg( {c_\alpha}^{\frac{1}{2}-k}  { \alpha^{k} K_{k-\frac{1}{2}}(c_\alpha)} \cdot \tau_2(0) + \sum\limits_{j=1}^\infty {c_j}^{\frac{1}{2}-k}  { j^{k}  K_{k-\frac{1}{2}}(c_j)} \cdot \tau_2(j) \bigg) & \text{if $k\geq1$}  \label{PIGtau4}
  \end{cases}
\end{align}
\section{The $\alpha^*$ and $\sigma^*$ for which $\tau_1(0)=\tau_2(0)$ (same proportion of zeros in the original and synthetic data)}
The parameter $\alpha^*$ can be set so that, for a given $\sigma^*$, the expected proportion of zeros in the synthetic data is equal to the proportion in the original data, that is, for $\tau_1(0)=\tau_2(0)$. 
\subsubsection{The negative binomial (NBI)}
\begin{align}
\tau_1(0)&=\tau_2(0) \quad \text{and from (\ref{tau1NBI})} \nonumber \\
\tau_2(0)&= \frac{1}{(1+\sigma^* \alpha^*)^{1/\sigma^*}} \cdot \tau_2(0) + \sum\limits_{j=1}^\infty \frac{1}{(1+\sigma^* j)^{1/\sigma^*}} \cdot \tau_2(j)\nonumber \\
\Longleftrightarrow (1+\sigma^* \alpha^*)^{1/\sigma^*} &= \bigg( 1-\frac{1}{\tau_2(0)}\sum\limits_{j=1}^\infty \frac{1}{(1+\sigma^* j)^{1/\sigma^*}} \cdot \tau_2(j)  \bigg)^{-1}\nonumber \\
\Longleftrightarrow \alpha^* &= \frac{1}{\sigma^*} \bigg[\bigg( 1-\frac{1}{\tau_2(0)}\sum\limits_{j=1}^\infty \frac{1}{(1+\sigma^* j)^{1/\sigma^*}} \cdot \tau_2(j)  \bigg)^{-\sigma^*} -1\bigg]. \nonumber
\intertext{Hence the required value, $\alpha^*$, can be derived analytically.
\subsubsection{The Poisson-inverse Gaussian (PIG)}}
\tau_1(0)&=\tau_2(0) \quad \text{and from (\ref{tau0PIG})} \nonumber \\
\tau_2(0)&=\text{exp}({1/\sigma^*}-c_\alpha^*) \cdot \tau_2(0) + \sum\limits_{j=1}^\infty{ \text{exp}({1/\sigma^*}-c_j)} \cdot \tau_2(j) \nonumber \\
\Longleftrightarrow c_\alpha^* &= \frac{1}{\sigma^*} - \text{log}\bigg(1-\frac{1}{\tau_1(0)} \sum\limits_{j=1}^\infty{ \text{exp}({1/\sigma^*}-c_j)} \cdot \tau_2(j) \bigg) \nonumber\\
\Longleftrightarrow \alpha^* &= \frac{1}{2} \bigg\{ \sigma^*\bigg[ \frac{1}{\sigma^*} - \text{log}\bigg(1-\frac{1}{\tau_1(0)} \sum\limits_{j=1}^\infty{ \text{exp}({1/\sigma^*}-c_j)} \cdot \tau_2(j) \bigg)\bigg]^2 -\frac{1}{\sigma^*} \bigg\}. \nonumber
\end{align}
\section{The $(\sigma^*,\alpha^*)$ for which $\tau_4(1)=p$} Alternatively, $\alpha$ and $\sigma$ can be adjusted such that $\tau_4(1)=p$ for some proportion $p$.
\subsubsection{The negative binomial (NBI)}
\begin{align}
\tau_4(1)&=p \quad \text{and setting $k=1$ in (\ref{NBItau4}),} \nonumber \\
p&=  \frac{\tau_2(1)}{(1+\sigma^*)^{1+1/\sigma^*}} \bigg/\bigg( \frac{\alpha^* \cdot \tau_2(0) }{(1+\alpha^*\sigma^*)^{1+1/\sigma^*}} + \sum\limits_{j=1}^\infty \frac{j \cdot \tau_2(j) }{(1+j\sigma^*)^{1+1/\sigma^*}} \bigg) . \nonumber
\intertext{
\subsubsection{The Poisson-inverse Gaussian (PIG)} 
When $\lambda=1/2$, the Bessel function of the third kind has a closed formed expression, $K_{1/2}(t)=(\pi/2t)^{1/2}\text{exp}(-t)$, and setting $k=1$ in (\ref{PIGtau4}) gives,}
p&=\bigg[\frac{\text{exp}(-c_1)}{c_1} \cdot \tau_2(1) \bigg]\bigg/\bigg[\frac{\alpha^*\cdot \text{exp}(-c_{\alpha^*})}{c_{\alpha^*}} \cdot \tau_2(0) + \sum\limits_{j=1}^\infty \frac{j \cdot \text{exp}(-c_j)}{c_j} \cdot \tau_2(j) \bigg]. \nonumber
\end{align}